\documentclass[11pt]{article}
\usepackage[a4paper, left=20mm, right=20mm, top=25mm, bottom=35mm]{geometry}
\usepackage{bm,amssymb,amsfonts,amsmath}  
\usepackage{graphicx}

\newcommand{\lleq}[1]{\label{#1}}

\newcommand{\bme}{{\bm{e}}}
\newcommand{\bmn}{{\bm{n}}}
\newcommand{\bmp}{{\bm{p}}}
\newcommand{\bmq}{{\bm{q}}}
\newcommand{\bms}{{\bm{s}}}

\newcommand{\bmK}{{\bm{K}}}
\newcommand{\bmP}{{\bm{P}}}
\newcommand{\bmalpha}{{\bm{\alpha}}}
\newcommand{\bmlambda}{{\bm{\lambda}}}

\newcommand{\ff}[1]{{\underline{#1}}}

\newcommand{\sav}[1]{{\bigl\langle #1 \bigr\rangle}}
\newcommand{\savo}[2]{{\bigl\langle #1 \bigr\rangle_{\!#2}}}
\def\cond {\,|\,}
\def\var{\mathrm{var}}
\def\cov{\mathrm{cov}}

\def\qghpa{{Q_{\scriptscriptstyle a}}}  
\def\qghpm{{Q_{\scriptscriptstyle m}}}  

\def\azia{{\mathcal A}}
\def\np{{n}}  
\def\nsp{{n_{\scriptscriptstyle +}}} 
\def\smallB{{\scriptscriptstyle B}}
\def\smallI{{\scriptscriptstyle I}}
\def\smallN{{\scriptscriptstyle N}}

\newcommand{\savN}[1]{{\left\langle #1 \right\rangle}_{\!\!\scriptscriptstyle N}}

\newcommand{\savNb}[1]{{\left\langle #1 \right\rangle}_{\!\!\scriptscriptstyle Nb}}

\newcommand{\savNl}[1]{{\left\langle #1 \right\rangle}_{\!\!\!\!\scriptscriptstyle N}}
\newcommand{\savNal}[1]{{\left\langle #1 \right\rangle}_{\!\!\!\!\scriptscriptstyle Na}}
\newcommand{\savNbl}[1]{{\!\!\left\langle #1 \right\rangle}_{\!\!\!\!\scriptscriptstyle Nb}}
\newcommand{\savNcl}[1]{{\!\!\left\langle #1 \right\rangle}_{\!\!\!\!\scriptscriptstyle Nc}}

\def\samplei{{\mathcal S}}
   \def\nev{{{\mathcal E}}}

\def\sampleAB{{{\mathcal S}_{AB}}}

\def\sampleN{{{\mathcal S}_{\scriptscriptstyle N}}}
   \def\nevN{{{\mathcal E}_{\scriptscriptstyle N}}}
   
   \def\evrN{{{\mathcal R}_{\scriptscriptstyle N}}}
   \def\evrNp{{{\mathcal R}_{\scriptscriptstyle N}^\prime}}
   \def\evrNpp{{{\mathcal R}_{\scriptscriptstyle N^\prime}^\prime}}

\def\samplenpN{{{\mathcal S}_{\np {\scriptscriptstyle N}}}}
   \def\nevnpN{{{\mathcal E}_{\np {\scriptscriptstyle N}}}}
   \def\evrnpN{{{\mathcal R}_{\np {\scriptscriptstyle N}}}}

\def\hro{\hat\rho}

\def\rhomult{\rho^{\rm mult}}
\def\rhoref{\rho^{\rm ref}}

\newcommand{\rhon}[1]{\rho_{\scriptscriptstyle #1 N}}

\newcommand{\kappan}[1]{\kappa_{\scriptscriptstyle #1 N}}

\begin{document}

\begin{center}
  \textbf{\Large Internal cumulants for femtoscopy with fixed charged multiplicity}\\[12pt]
  H.C.\ Eggers$^a$ and B.\ Buschbeck$^b$\\
  $^a$\textit{Department of Physics, University of Stellenbosch, 7602
    Stellenbosch, South Africa} \\
  $^b$\textit{Institut f\"ur Hochenergiephysik, Nikolsdorfergasse 18,
    A--1050 Vienna, Austria}\\
\end{center}

\begin{abstract}
  A detailed understanding of all effects and influences on
  higher-order correlations is essential. %
  At low charged multiplicity, the effect of a nonpoissonian
  multiplicity distribution can significantly distort correlations. %
  Evidently, the reference samples with respect to which correlations
  are measured should yield a null result in the absence of
  correlations. %
  We show how the careful specification of desired properties
  necessarily leads to an average-of-multinomials reference sample.
  The resulting internal cumulants and their averaging over several
  multiplicities fulfil all requirements of correctly taking into
  account nonpoissonian multiplicity distributions as well as yielding
  a null result for uncorrelated fixed-$N$ samples.
  Various correction factors are shown to be approximations at
  best.
  Careful rederivation of statistical variances and covariances within
  the frequentist approach yields errors for cumulants that differ
  from those used so far.
  We finally briefly discuss the implementation of the analysis through
  a multiple event buffer algorithm.
  
\end{abstract}

\tableofcontents

\section{Introduction and motivation}
\label{sec:intro}

The understanding of hadronic collisions is now considered an
essential baseline for ultrarelativistic heavy-ion collisions.  Given
the correspondingly low final-state multiplicities, there are
significant deviations, even for inclusive samples, from assumptions
commonly made both in the general theory and in the definition of
experimentally measured quantities such as a nongaussian shape of the
correlation function and nonpoissonian multiplicity
distributions. Constraints such as energy-momentum conservation %
\cite{2008-ChajeckiLisa-prc78.064903-nuclth0803.0022,
  2009-ChajeckiLisa-prc79.034908-nuclth0807.3569} 
would also play a role in at least some regions of phase space.
Multiplicity-class and fixed-multiplicity analysis differ increasingly
from poissonian and inclusive distributions, and with the good
statistics now available, measurements have become accurate enough to
require proper understanding and treatment of these assumptions and
deviations, which play an ever larger role with increasing order of
correlation.

\subsection{Correlations as a function of charged multiplicity}
\label{sec:fxml}

There are a number of reasons to study correlations at fixed charged
multiplicity $N$ or, if necessary, charged-multiplicity classes. 
Firstly, the physics of multiparticle correlations will evidently
change with $N$, and indeed the multiplicity dependence of various
quantities such as the intercept parameter and radii associated with
Gaussian parametrisations is under constant scrutiny %
\cite{1987-ABCDHWBreakstone-zpc33.333-xx.xx,
2000-BuschbeckEggersLipa-plb481.187-hepex0003029,
2001-BuschbeckEggers-npbps92.235-hepph0011292,
2001-BuschbeckEggersN-npbps92.235-hepph0011292, 
2005-KittelDeWolf-SoftMultihadronDynamics,
2005-LisaPrattSoltzWiedemann-arnps55.357-nuclex0505014,
2009-Chajecki-appb40.1119-nuclex0901.4078,
2010-CMS-prl105.032001-hepex1005.3294,
2011-CMS-jhep05.029-hepex1101.3518,
2012-Alexander-jpg39.085007-hepph1202.3575}. %
Measurement of many observables as a function of multiplicity class,
regarded a proxy for centrality dependence, has been routine for
years.  Corresponding theoretical considerations, e.g.\ in the quantum
optical approach go back a long time
\cite{1999-SuzukiBiyajima-prc60.034903-hepph9907348}.
Secondly, correlations for fixed-$N$ are the building blocks which are
combined into multiplicity-class- and inclusive correlations
\cite{1973-Koba-appb4.95-xx.xx}.

However, such fixed-$N$ correlations have been beset by an
inconsistency in that they are nonzero even when the underlying sample
is uncorrelated and do not integrate to zero either.  This has been
recognised from the start \cite{1975-Foa-pr22.1-xx.xx}, and various
attempts have been made to fix the problem.

Combining events from several fixed-$N$ subsamples into multiplicity
classes does not solve these problems. To quote an early reference
\cite{1975-Eggert-npb86.201-xx.xx}: %
\textit{Averaging over multiplicities inextricably mixes the
  properties of the correlation mechanism with those of the
  multiplicity distribution. Instead, the study of correlations at
  fixed multiplicities allows one to separate both effects and to
  investigate the behaviour of correlation functions as a function of
  multiplicity.} %
Under the somewhat inappropriate name of ``Long-Range Short-Range
correlations'' \cite{1975-Foa-pr22.1-xx.xx,%
  1991-Carruthers-pra43.2632-xx.xx}, 
an attempt was made to separate these multiplicity-mixing correlations
from the fixed-$N$ correlations, but the inconsistencies inherent in
the underlying fixed-$N$ correlations were not addressed. Building on
Ref.~\cite{1996-LipaEggersBuschbeck-prd53.4711-hepph9604373}, we
propose doing so now.

\subsection{Cumulants in multiparticle physics}

Multiparticle cumulants have entered the mainstream of analysis, as
shown by the following incomplete list of topics.  In principle, the
considerations presented in this paper would apply to any and all such
cumulants to the degree that their reference distribution deviates
from a poisson process or that the type of particle kept fixed differs
from the particle being analysed.

  Integrated cumulants of multiplicity distributions have a long
  history in multiparticle physics \cite{1971-Mueller-prd4.150}. %
  Second-order differential cumulants, normally termed ``correlation
  functions'', have likewise been ubiquitous for decades
  \cite{2005-KittelDeWolf-SoftMultihadronDynamics} both in
  charged-particles correlations \cite{1975-Foa-pr22.1-xx.xx} and in
  femtoscopy since they provide information on spacetime
  characteristics of the emitting sources, %
  most recently at the LHC %
  \cite{2010-CMS-prl105.032001-hepex1005.3294,
    2011-CMS-jhep05.029-hepex1101.3518,
    2010-ALICE-prd82.052001-nuclex1012.4035}.
  Differential three-particle cumulants generically measure
  asymmetries in source geometry and exchange amplitude phases
  \cite{1997-HeinzZhang-prc56.426-nuclth9701023}. %
  They also provide consistency checks
  \cite{1997-EggersLipaBuschbeck-prl79.197-hepph9702235} %
  and a tool to disentangle the coherence parameter from other effects
  \cite{1993-AndreevPlumerWeiner-ijmpa8.4577-xx.xx,
    2008-CsorgoKittelMetzgerNovak-plb663.214-hepph0803.3528}. 
  Three-particle cumulants are also sensitive to differences between
  longitudinal and transverse correlation lengths in the Lund model
  \cite{1998-AnderssonRingner-plb421.283-hepph9710334}.
%
  Inclusive three-particle cumulants have been measured,
  albeit with different methodologies, %
  in, for example, hadronic
  \cite{1978-Kenney-npb144.3112-xx.xx,
  1992-UA1Neumeister-plb275.186-xx.xx,
  1994-NA22Agababyan-plb332.458-xx.xx,
  1995-NA22Agababyan-zpc68.229-xx.xx}, 
  leptonic  %
  \cite{1995-DELPHIAbreu-plb355.415-xx.xx,
    1998-OPALAckerstaff-epjc5.239-hepex9806036,
    2002-L3-plb540.182-hepex0206051,
    2001-OPALAbbiendi-plb523.35-hepex0110051} 
  and nuclear collisions
  \cite{2000-WA98-prl85.2895-hepex0008018,
    2003-STAR-prl91.262301-nuclex0306028,
    1998-NA44Sakaguchi-npa638.103c-xx.xx,
    2001-NA44-plb517.25-nuclex0102013}. 
  They play a central role in direct QCD-based calculations %
  \cite{1997-KhozeOchs-ijmpa12.2949-hepph9701421,
    1999-BuschbeckMandl-plb457.368-hepph9905367,
    2011-PerezMathieuSanchis-prd84.034015-hepph1104.1973} 
  and in some recent theory and experiment of azimuthal and
  jet-like correlations %
  \cite{2006-Pruneau-prc74.064910-nuclex0608002,
    2008-UleryWang-nima595.502-nuclex0609016,
    2009-Pruneau-prc79.044907-nuclex0810.1461,
    2012-ATLAS-jhep1207.019-hepex1203.3100,
    2012-STARYi-xx.xx-nuclex1210.6640}. 
  Net-charge and other charge combinations are considered
  probes of the QCD phase diagram %
  \cite{2011-Pratt-prc85.014904-nuclth1109.3647,
    2012-BzdakKoch-xx.xx-nuclth1206.4286,
    2012-ALICE-xx.xx-nuclex1207.0900}. %
  Cumulants of order 4 or higher are, of course, increasingly
  difficult to measure and so early investigations were largely
  confined to their scale dependence %
  \cite{1991-CarruthersEggersSarcevic-plb254.258-xx.xx,
    1996-DewolfDreminKittel-pr270.1-hepph9508325,
    2000-Sarkisyan-plb477.1-hepph0001262,
    2000-AlexanderSarkisyan-plb487.215-hepph0005212}. 
  The large event samples now available have, however, made feasible
  measurements of fourth- and higher-order cumulants in other
  variables as proposed in
  \cite{1993-EggersLipaCarruthersBuschbeck-plb301.298-xx.xx,
    1996-CramerKadija-prc53.908-xx.xx,
    1999-SuzukiBiyajima-prc60.034903-hepph9907348,
    1999-CsorgoLorstadSchmidtSter-epjc9.275-hepph9812422,
    2001-BorghiniDinhOllitrault-prc64.054901-nuclth0105040} 
  as, for example, recently measured by ALICE
  \cite{2010-ALICE-prl105.252302-xx.xx}.
  Reviews of femtoscopy theory range from
  \cite{1989-Podgoretsky-sjpn20.266,
    1996-AmelinLednicky-hip4.241,
    1999-HeinzJacak-arnps49.529-nuclth9902020} 
  to more recent ones such as %
  \cite{2005-LisaPrattSoltzWiedemann-arnps55.357-nuclex0505014}.

\subsection{Outline of this paper}

It has long been obvious that the root cause of the problems and
inconsistencies set out in Section \ref{sec:fxml} was the reference
sample \cite{2001-Eggers-Tihany-hepex0102005}. Insofar as cumulants
are concerned, the solution was outlined in
Ref.~\cite{1996-LipaEggersBuschbeck-prd53.4711-hepph9604373} as a
subtraction of the reference sample cumulant from the measured one;
important pieces of the puzzle were, however, still missing at that
stage.  In this paper, we clarify and extend the basic concept of
internal cumulants and consider in detail the case of second- and
third-order differential cumulants in the invariant $Q =
\sqrt{-(p_1-p_2)^2}$ for fixed charged multiplicity $N$. The method
may be implemented for other variables without much fuss.

A second cornerstone of the present paper is the recognition that the
$n$ particles which enter a correlation analysis are usually only a
subset of the $N$ charged pions. While in the case of charged-particle
correlations all $N$ particles are used in the analysis, Bose-Einstein
correlations, for example, would use only the $n \equiv n_+$ positive
pions (and, in a separate analysis, only the $n_- = N-n_+$ negatives).
In addition, there may be reasons to restrict the analysis itself to
subregions of the total acceptance $\Omega$ in which $N$ was measured,
as exemplified in this paper by restriction to a ``good azimuthal
region'' subinterval around the beam axis, $\azia \subset [0,2\pi]$,
in which detection efficiency is high. $\azia$ can, however, be
reinterpreted generically as any restriction in momentum space
compared to $\Omega$ and/or as a selection such as charge or particle
species. Even when setting $\azia=\Omega$ i.e.\ doing the femtoscopy
analysis in the full acceptance $n$ still does not equal $N$ but
fluctuates around $N/2$. The trivial observation that $n\neq N$
fundamentally changes the analysis: \textit{identical-particle
  correlations at fixed $N$ and charged-particle correlations at fixed
  $N$ require different definitions.}  %

As we shall show, ad hoc prescriptions such as simply inserting
prefactors or implementing event mixing using only events of the same
$N$ do alleviate the effect of the overall nonpoissonian multiplicity
distribution in part but fail to remove them completely.  The same
issues will, of course, arise in any other correlation type of, for
example, nonidentical particles or net charge correlations. The
formalism set out here can be easily extended to such cases.  A
refined version of the abovementioned Long-Range-Short-Range method,
which we term ``Averaged-Internal'' cumulants, will be presented in
Section \ref{sec:smpab}.  Along the way, we document in Section
\ref{sec:rdct} extended versions of the particle counters
\cite{1967-Klimontovich,1963-BroutCarruthers-ManyElectron} which we
need as the basis for correlation studies and in Section
\ref{sec:stderrors} demonstrate from first principles that statistical
errors for cumulants used so far have captured only some of the terms
and with partly incorrect prefactors. Section \ref{sec:vtmx} outlines
the implementation of event mixing for fixed-$N$ analysis.  While
experimental results will be published elsewhere, preliminary results
in Figs.~2 and 3 show that, in third and even in second order,
corrections due to proper treatment of fixed-$N$ reference samples can
be large.

\section{Raw data, counters and densities}
\label{sec:rdct}

\subsection{Raw data}

The starting point for experimental correlation analysis is the
inclusive sample $\samplei$, made up of $\nev$ events $a =
1,\ldots,\nev$. Each event consists of a varying number of final-state
elementary particles and photons; for our purposes, we consider only
the $N(a)$ charged pions of event $a$ in $\Omega$, the maximal
acceptance region used.  Each pion $i = 1,\ldots,N(a)$ is
characterised by a data vector $(\bmP_i^a,\bms_i^a,\bme_i^a)$
containing its measured information, including the three components of
its momentum, $ \bmP_i^a = (p_{ix}^a,p_{iy}^a,p_{iz}^a) \text{ or }
(y_i^a,\phi_i^a,p_{ti}^a)$, while its discrete attributes such as
mass, charge etc.\ are captured in a data vector $\bms_i^a$ of
discrete values; for the moment, we shall consider only the charge,
$\bms_i^a\to c_i^a$.  From the sample's raw data, we can immediately
find derived quantities such as the total charged multiplicity $N(a)$,
the total transverse energy etc., and such derived quantities are
hence considered part of the raw data.  The list of particle
attributes should be augmented by an error vector $\bme_i^a$
containing the measurement errors for each track, but we shall not
consider detector resolution errors here. In summary, the inclusive
data sample is fully described in terms of $\samplei =
\{\bmP^a,\bm{s}^a\}_{a=1}^\nev$ consisting of lists of vectors in
continuous and discrete spaces
\begin{align}
  \lleq{spe}
  \bmP^a &= \{\bmP_1^a,\bmP_2^a,\ldots,\bmP_{N(a)}^a\}
  && \bm{s}^a = \{\bm{s}_1^a,\bm{s}_2^a,\ldots,\bm{s}_{N(a)}^a\}.
\end{align}

\subsection{Data after conditioning and cuts}
\label{sec:ua1s}

For a particular analysis, the inclusive sample is invariably
subdivided and modified through ``conditioning'', the statistics
terminology for semi-inclusive or triggered analysis: From the total
sample of events, a subsample is selected according to some
restriction or precondition. In our case, this conditioning proceeds
in the following steps:
\begin{itemize}

\item \textit{Conditioning into fixed-$N$ subsamples:} For the
  fixed-multiplicity analyses that form the subject of this paper,
  $\samplei$ is subdivided into a set of fixed-$N$ subsamples
  $\sampleN$, each of which contains only events $a$ whose measured
  multiplicity $N(a)$ is equal to the constant $N$ characterising
  $\sampleN = \{ \bmP^a,\bm{s}^a \cond \delta(N,N(a))\}, \; N =
  0,1,2,\ldots$ We use the vertical bar $|$ here and everywhere in the
  usual sense of ``conditioning'' whereby the events in sample
  $\sampleN$ must satisfy the condition that their charged
  multiplicity must equal the specified constant $N$, denoted in this
  case by the Kronecker delta $\delta(N,N(a))$.  Quantities to the
  right of the vertical bar are generally considered known and fixed,
  while quantities left of the bar are variable or unknown.  The
  number of events in $\sampleN$ equals the $\delta$-restricted sum
  over the inclusive sample,
  \begin{align}
    \lleq{dtc}
    \nevN &= \sum_{a=1}^\nev \delta(N,N(a)),
    \qquad\qquad
    \sum_{N=0}^\infty \nevN = \nev.
  \end{align}
  The usual multiplicity distribution is the list of relative
  frequencies,\footnote{These event ratios
    are not probabilities in the strict sense; in the frequentist
    definition of probability, the two are equal only in the limit
    $\nev\to\infty$. We therefore avoid the use of the symbol
    $P_\smallN$ for such and similar data ratios.}
  \begin{align}
    \lleq{dtd}
    \evrN &= \frac{\nevN}{\nev}, \qquad\qquad
    \sum_{N=0}^\infty \evrN = 1.
  \end{align}
  While desirable, it is not easy to measure the total multiplicity of
  final-state charged pions, a quantity which approximately tracks the
  variation in the physics. Choosing charged pions measured within the
  maximal detector acceptance $N = N(\Omega)$ as marker is in any case
  only an approximation because it excludes charged particles outside
  the primary cuts and also ignores final-state particles other than
  charged pions. Nevertheless, we expect $N$ to be a reasonable
  measure of the multiplicity dependence of the
  physics. Alternatively, the multiplicity density in pseudorapidity
  at central rapidities $dN/d\eta$ can be used as a model-dependent proxy for $N$.

\item \textit{Azimuthal cut:} While $N(a)$ is the charged multiplicity
  measured in $\Omega$, there is no \textit{a priori} reason why the
  correlation analysis itself may not be conducted within a restricted
  part $\azia \subset \Omega$ of momentum space within which the
  actual analysis is done. In the case of the UA1 detector from which
  the data used in the examples below was drawn, $\azia$ refers to
  azimuthal regions within which measurement efficiency was high,
  and pions found in the low-efficiency azimuthal regions were excluded.
  Correspondingly, the multiplicity $n(a)$ which enters the
  correlation analysis itself differs from $N(a)$ and will, for a
  given fixed $N$ fluctuate with relative frequency
  \begin{align}
    \lleq{dtf} \evrnpN = \frac{\nevnpN}{\nevN} \qquad \text{for each
      fixed } N = 1, 2, \ldots ,
  \end{align}
  where $\nevnpN$ is the number of events for which $N(a)=N$ and $n(a)=n$.
  The outcome space for $n(a)$ will depend on its definition; in the
  present case where only positive (or only negative) pions within
  $\azia$ are used in the analysis, it will be $[0,1,\ldots,N]$
  so that the relative frequency is normalised by
  \begin{align}
    \lleq{dtg}
    \sum_{n=0}^N \evrnpN = 1 \qquad \forall\; N = 1, 2, \ldots .
  \end{align}
  With approximate charge conservation $n_+ \simeq n_-$, we expect
  the fixed-$N$ average for positive (or negative) pions in $\azia$ to
  hover around
  \begin{align}
    \lleq{dth}
    \langle n \rangle_{\!\smallN} 
    &\simeq \frac{N}{2}\,\frac{\text{volume of }\azia}{\text{volume of }\Omega}.
  \end{align}
  An example of the resulting relative frequencies (conditional
  normalised multiplicity distributions) is shown in Fig.~1. Since $n
  \leq N$, these conditional multiplicity distributions are almost
  always subpoissonian, i.e.\ narrower than a Poisson distribution
  with the same $\savN{n}$ would be.
  \begin{figure}
    \centerline{\includegraphics[width=110mm]{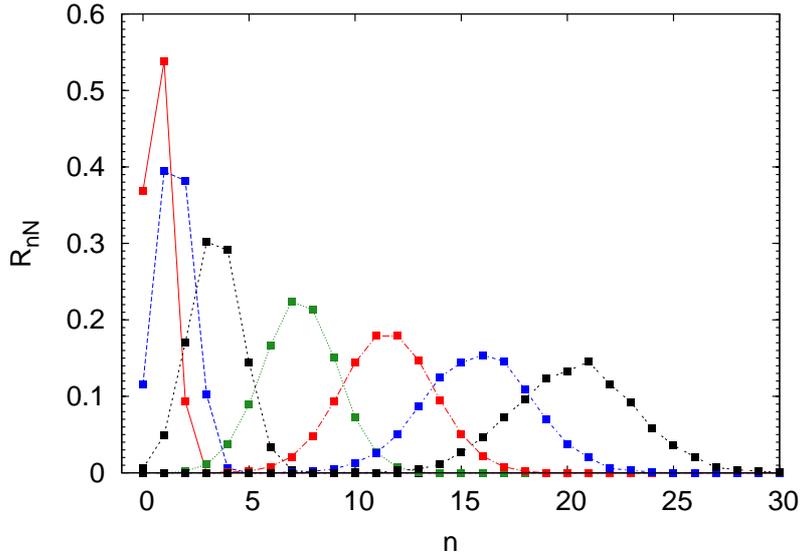}}
    \caption{\small Conditional normalised multiplicity distribution
      (relative frequency) of the number of positive pions $n$ in
      restricted azimuthal region $\azia = \{20^\circ \leq |\phi| \leq
      160^\circ\}$ for respectively fixed charged multiplicity $N =$ 3, 5, 10, 20,
      30, 40, 50, for the UA1 dataset used in
      Ref.~\cite{2006-EggersOctoberBuschbeck-plb635.280-hepex0601039}. In
      accordance with Eq.~(\ref{dth}), $\savN{n} \simeq 0.39 N$.}
  \end{figure}
  
\item \textit{Generalisation:} While in this paper the analysis will
  be carried out for the $n(a)$ positive pions of event $a$ falling
  into $\azia$, the same formalism obviously applies to negative pions
  and may equally refer to any other particles such as kaons, baryons,
  photons etc in any combination which depends on $N$.  There is no
  \textit{a priori} connection between the definitions of $N$ and $n$.

\item \textit{Identical-particle vs multi-species analysis:} While we
  do not develop the formalism for correlations between two or three
  particles of different species or charge, the methodology developed
  here can be easily modified to deal with such cases. For example,
  positive-negative pion combinations and ``charge balance
  correlations''
  \cite{2011-SchlichtingPratt-prc83.014913-nuclth1009.4283} can be
  handled by inserting delta functions $\delta(c,c_i^a)$, where $c$ is
  the desired charge and $c_i^a$ the measured charge of track $i$ in
  event $a$, into the definitions of the counters in Section
  \ref{sec:cntrs}.

\end{itemize}

\subsection{Counters and densities for fixed $N$}
\label{sec:cntrs}

This section is based on an old formalism
\cite{1967-Klimontovich,1963-BroutCarruthers-ManyElectron,
  1993-EggersLipaCarruthersBuschbeck-plb301.298-xx.xx} which must,
however, be updated to accommodate the issues being considered here.
The basic building block of correlation analysis is the
\textit{counter}; it is a particular projection of the raw data
particular suited to the construction of histograms.  Eventwise
counters $\hro$ for a given event $a$ are averaged to give sample
counters $\rho$.

We take the simple case where event $a$ contains $N(a)$ tracks with
three-momenta $\bmP^a = \{\bmP_1^a,\ldots,\bmP_{N(a)}^a\}$, no
discrete attributes $\bms$ and no further cuts or selection. For each
point $\bmp_1$ in momentum space, only that particle $i$ (if any)
whose momentum $\bmP_i^a$ happens to coincide with $\bmp$ is to be
counted,
\begin{equation} \lleq{ntc}
  \hro(\bmp_1\cond \bmP^a) = \sum_{i=1}^{N(a)} \delta(\bmp_1 - \bmP_i^a).
\end{equation}
Such counters always appear under an integral over some region of the
$\bmP$-space, so that the delta functions fulfil the purpose of
counting those particles falling within that region. Alternatively,
one can consider the delta functions here and below to represent small
nonoverlapping intervals around the specified momenta. The integral
over the full momentum space $\Omega$ yields
\begin{align} \lleq{ntd}
  \int_\Omega d\bmp_1\,\hro(\bmp_1\cond \bmP^a ) = N(a)
\end{align}
while an integral over some subspace or bin $\Omega_b \subset \Omega$
will yield the number of particles of event $a$ in bin $\Omega_b$.
The second-order eventwise counter for event $a$ is\footnote{Pairs are
  \textit{ordered}, i.e.\ a particular pair is counted
  twice. Unordered pair counting is possible but unnecessarily
  complicates sum limits.}
\begin{equation} \lleq{nte}
  \hro(\bmp_1,\bmp_2\cond \bmP^a) 
  = \sum_{i\ne j = 1}^{N(a)} \delta(\bmp_1-\bmP_i^a)\, \delta(\bmp_2-\bmP_j^a),
\end{equation}
with the inequality $i\ne j$ ensuring that a single particle is not
counted as a ``pair''. The counter integrates to
\begin{align} \lleq{ntf}
  \int_\Omega d\bmp_1\,d\bmp_2\,\hro(\bmp_1,\bmp_2\cond \bmP^a) 
  &= N(a)(N(a) -1) \ =\  N(a)^{\ff{2}}
\end{align}
using the falling factorial notation
\begin{align} \lleq{ntg}
  N^{\ff{r}} = N(N-1)\cdots(N-r+1) \qquad r = 1,2,\ldots
\end{align}
as contrasted to the rising factorial (Pochhammer symbol)
$N^{\overline{r}} = N(N+1)\cdots(N+r-1)$.  The single-particle counter
is a projection of $\hro(\bmp_1,\bmp_2\cond \bmP^a)$ because
\begin{equation} \lleq{nti}
  \int_\Omega d\bmp_2\, \hro(\bmp_1,\bmp_2\cond \bmP^a) 
  = [N(a)-1]\,\hro(\bmp_1\cond \bmP^a).
\end{equation}
The most general eventwise counter which enters the exclusive cross
section for events with charged multiplicity $N(a)$
\begin{align}\lleq{ntj}
  \hro(\bmp_1,\bmp_2,\ldots,\bmp_{N(a)}\cond \bmP^a)
  = \sum_{i_1\ne i_2\ne \cdots \ne i_{N(a)} = 1}^{N(a)}
  \;\prod_{d=1}^{N(a)} \delta(\bmp_d - \bmP_{i_d}^a)
\end{align}
fully describes the event, including any and all correlations between
its particles. It integrates to the factorial of the event
multiplicity
\begin{align}\lleq{ntk}
  \int_\Omega d\bmp_1\,d\bmp_2 \,\cdots\,d\bmp_{N(a)}\;
  \hro(\bmp_1,\bmp_2,\ldots,\bmp_{N(a)}\cond \bmP^a)
  = N(a)!
\end{align}
and contains all counters of lower order by projection. An $r$th-order
counter $\hro(\bmp_1,\bmp_2,\ldots,\bmp_r\cond \bmP^a)$ is zero
whenever there are more observation points than particles being
observed, $r > N(a)$.\footnote{A counter of order $N(a)$ can be made
  to behave like one of order $N(a)+1$ by defining an additional dummy
  data point $\bmP_{N(a)+1}$ which lies \textit{outside} the normal
  domain $\Omega$, but we shall not pursue this here..}  %

To distinguish eventwise counters for nonfixed $N$ from eventwise
counters for fixed $N$, we define the separate eventwise
counter for fixed $N$ by specifying an additional Kronecker delta,
\begin{align}
  \lleq{ntl}
  \hro(\bmp_1,\ldots,\bmp_r\cond \bmP^a,N) &= \delta(N,N(a))
  \sum_{i_1\ne \cdots\ne i_r}^N \prod_d \delta(\bmp_d - \bmP_{i_d}^a)
  \qquad r = 1,2,\ldots,N.
\end{align}

While the counters and densities defined above and below are clearly
frame-dependent, it is easy to define corresponding Lorentz-invariant
versions by supplementing each delta function in 3-momenta with the
corresponding energy; thus Eq.~(\ref{ntc}) would become, for example,
\begin{equation} \lleq{ntci}
  \hro(\bmp_1\cond \bmP^a) = \sum_{i=1}^{N(a)} E(\bmp_1)\, \delta(\bmp_1 - \bmP_i^a)
\end{equation}
with $E(\bmp_1)= E_1 = \sqrt{\bmp_1^2+m^2}$ the on-shell energy, and in
general
\begin{align}
  \lleq{ntji}
  \hro(\bmp_1,\bmp_2,\ldots,\bmp_r\cond \bmP^a)
  &= \qquad \sum_{i_1\ne i_2\ne \cdots \ne i_r = 1}^{N(a)}
  \;\prod_{d=1}^{N(a)} E_d\,\delta(\bmp_d - \bmP_{i_d}^a) 
  \quad r = 1,\ldots,N(a), \\
  \lleq{ntni}
  \hro(\bmp_1,\ldots,\bmp_r\cond \bmP^a,N) &= \delta(N,N(a))
  \sum_{i_1\ne \cdots\ne i_r}^N \prod_d E_d\,\delta(\bmp_d - \bmP_{i_d}^a)
  \quad r = 1,\ldots,N,
\end{align}
which are manifestly invariant. Because such counters and densities
are, however, always integrated over some $\Omega_b$ by $\prod_d
(d\bmp_d/E_d)$, the additional factors $E_d$ always cancel and play no
role on this level of analysis and will be ignored for the time
being. The bin boundaries of $\Omega_b$ do, however, remain
frame-dependent.

Charge-, spin- or species-specific counters are defined in the same
way, i.e.\ by supplying appropriate Kronecker deltas to the counters;
for example the particle counter for pions with charge $c_1$ at
$\bmp_1$ for fixed $N$ is
\begin{align}
  \lleq{ntnb}
  \hro(\bmp_1\cond c_1,\bmP^a,N)
  &= \delta(N,N(a))\, \sum_{i=1}^{N(a)}   \delta(c_1,c_i^a)\,\delta(\bmp_1 - \bmP_i^a)\,
\end{align}
while the two-particle counter for charge combination $(c_1,c_2)$ at
momenta $(\bmp_1,\bmp_2)$ for any $N$ is, for example,
\begin{align}
    \lleq{nto}
  \hro(\bmp_1,\bmp_2\cond c_1,c_2,\bmP^a) 
  &= \sum_{i,j=1}^{N(a)} 
  \delta(c_1,c_i^a)\,\delta(\bmp_1 - \bmP_i^a)\,
  \delta(c_2,c_j^a)\,\delta(\bmp_2 - \bmP_j^a) .
\end{align}
In contrast to Eq.~(\ref{ntnb}), charge counters rather than particle
counters would be
\begin{align}
  \lleq{ntq}
  \hro_c(\bmp_1\cond c_1,\bmP^a)
  &= \sum_{i=1}^{N(a)} c_1\, \delta(c_1,c_i^a)\,\delta(\bmp_1 - \bmP_i^a)\,
\end{align}
so that $\hro_c(\bmp_1\cond {+}1,\bmP^a) + \hro_c(\bmp_1\cond
{-}1,\bmP^a)$ represents the net charge of event $a$ at $\bmp_1$.  The
two-particle counter for charges $(c_1,c_2)$ at momenta
$(\bmp_1,\bmp_2)$ is
\begin{align}
    \lleq{ntr}
    \hro^{c_1\,c_2}  \ \equiv\
  \hro_c(\bmp_1,\bmp_2\cond c_1,c_2,\bmP^a) 
  &= \sum_{i,j=1}^{N(a)} 
  c_1\,\delta(c_1,c_i^a)\,\delta(\bmp_1 - \bmP_i^a)\,
  c_2\,\delta(c_2,c_j^a)\,\delta(\bmp_2 - \bmP_j^a),
\end{align}
and ``charge flow'' correlations can be constructed from this (for
rapidities $(y,y')$ in the case of Ref.\ \cite{1981-TASSO-plb100.357}) 
such as
\begin{align}
  \lleq{nts}
  \Phi(y,y') 
  &= - \langle \textstyle 
  \sum_{i\neq j} c_i^a \,c_j^a \,\delta(y - Y_i^a)\,\delta(y' - Y_j^a) \rangle
\end{align}
which can be expressed as $\Phi = \langle \hro^{+-} + \hro^{+-} -
\hro^{++} - \hro^{--} \rangle$ and the related ``charge balance
functions'' described in e.g.\
\cite{2011-Pratt-prc85.014904-nuclth1109.3647}.

Returning to the fixed-$N$ case: eventwise counters will usually be
combined with similar events to form event averages.  The simplest
average is the fixed-$N$ density for the subsample of fixed
$\sampleN$,
\begin{align}\lleq{vcd}
  \rho(\bmp_1\cond \sampleN)
  &= \frac{1}{\nevN} \sum_a \hro(\bmp_1\cond \bmP^a,N),
\end{align}
with the Kronecker delta in (\ref{ntl}) ensuring that only events in
$\sampleN$ are considered, so we need not further specify the
individual terms or limits of the $a$-sum. Using (\ref{dtc}), it is
immediately clear that
\begin{align}\lleq{vcf}
  \int_\Omega d\bmp_1\,  \rho(\bmp_1\cond \sampleN)
  &= N
\end{align}
compared to the integral over the corresponding eventwise counter
\begin{align}\lleq{vcg}
  \int_\Omega d\bmp\,  \hro(\bmp\cond \bmP^a,N)
  &= N\,\delta(N,N(a))
\end{align}
and to the integral (\ref{ntd}); similarly
\begin{align}
  \lleq{vch}
  \int_\Omega d\bmp_1\cdots d\bmp_r\,
  \rho(\bmp_1,\ldots,\bmp_r\cond \sampleN) 
  = N(N{-}1)\cdots(N{-}r+1) = N^{\ff{r}}.
\end{align}
The inclusive averaged density $\rho(\bmp_1\cond \samplei)$ is the
weighted average over all $N$ of the fixed-$N$ averages,
\begin{equation}
  \lleq{vci}
  \rho(\bmp_1\cond  \samplei) 
  = \sum_{N=1}^\infty \evrN\, \rho(\bmp_1\cond \sampleN).
\end{equation}
Using (\ref{dtd}),(\ref{ntl}) and (\ref{vcd}), this can be written as
\begin{equation}\lleq{vcj}
  \rho(\bmp_1\cond \samplei) 
  = \frac{1}{\nev} \sum_a \sum_i \delta(\bmp_1-\bmP_i^a),
\end{equation}
and for general $r=1,2,\ldots$
\begin{equation}\lleq{vck}
  \rho(\bmp_1,\ldots,\bmp_r\cond\samplei) 
  = \sum_{N=r}^\infty \evrN\, \rho(\bmp_1,\ldots,\bmp_r\cond \sampleN)
  = \frac{1}{\nev} \sum_a \sum_{i_1\neq\cdots\neq i_r} \prod_{d=1}^r \delta(\bmp_d-\bmP_{i_d}^a),
\end{equation}
keeping in mind that $\hro(\bmp_1,\ldots,\bmp_r\cond\sampleN)$ will be
zero whenever $N(a) < r$. The integral of any $r$th order inclusive
averaged density is the $r$th-order factorial moment of the
multiplicity distribution,
\begin{align}
  \lleq{vcl}
  \int_\Omega d\bmp_1 \cdots d\bmp_r\,
  \rho(\bmp_1,\ldots,\bmp_r\cond \samplei) 
  = \sum_{N=r}^\infty \evrN\, N^{\ff{r}}
  = \left\langle N^{\ff{r}} \right\rangle
\end{align}
with simple angle brackets denoting inclusive averaging.

The averaged counters are of course directly related to the
traditional definitions in terms of cross sections. If ${\cal L}$ is
the integrated luminosity of incoming particles, the topological cross section is
$\sigma_\smallN = \nevN/{\cal L}$, the inelastic cross section is
$\sigma_\smallI = \nev/{\cal L}$ and the inclusive cross section is
$\sigma_{\rm incl} = \sum_\smallN N\sigma_\smallN = \langle N \rangle
\,\sigma_\smallI$ while the relative frequency (multiplicity
distribution) can be written as usual as $\evrN =
\sigma_\smallN/\sigma_\smallI$.  The relation between the differential
cross sections and our counters is
\begin{align}
    \rho_{\rm incl} (\bmp_1,\ldots,\bmp_\smallN)
    \ = \
    \rho(\bmp_1,\ldots,\bmp_\smallN\cond \samplei)
    &= \frac{1}{\sigma_\smallI} \frac{d^{3N}\sigma_{\rm incl}}{d\bmp_1 \cdots d\bmp_\smallN}, \\
    \rho(\bmp_1,\ldots,\bmp_\smallN\cond \sampleN) 
    &= \frac{1}{\sigma_\smallN} \frac{d^{3N}\sigma_{\rm excl}}{d\bmp_1 \cdots d\bmp_\smallN}, 
\end{align}
and so as usual inclusive and exclusive densities are related by
\cite{1973-Koba-appb4.95-xx.xx}
\begin{equation}
    \rho(\bmp_1,\ldots,\bmp_\smallN\cond \samplei)
  = \sum_{N=r}^\infty \frac{\evrN}{(N-r)!} \int 
  \rho(\bmp_1,\ldots,\bmp_\smallN\cond \sampleN) \,d\bmp_{r+1}\cdots d\bmp_\smallN,
\end{equation}
while the semi-inclusive cross sections and counters follow by the
usual projections.

\subsection{Counters and densities for fixed $(N,\np)$}

Our choice of a basic counter is motivated by the experimental
situation set out in Section \ref{sec:intro}: we wish to work in event
subsamples of fixed total charged multiplicity $N(a)$ in the entire
momentum space $\Omega$, but do the differential correlation analysis
using only those pions $\np$ which fall into the restricted space
$\azia$ and of a particular charge $+1$ or $-1$. This requires the use
of ``subsubsamples'' for which both $N$ and $\np$ are kept fixed,
\begin{align}
  \samplenpN &= \bigl\{ \bmP^a, \text{ with $a$ constrained by }
  \delta(\np,\nsp(a))\;\delta(N,N(a)) \bigr\}
\end{align}
with $\nsp(a)$ the number of positive pions of event $a$ in $\azia$,
and eventwise subsubsample counters
\begin{align}
  \lleq{vdg}
  \hro(\bmp_1,\ldots,\bmp_r\cond \np,N,\bmP^a)
  &= \delta(\np,\nsp(a))\, \delta(N , N(a))\, \sum_{i_1\ne \cdots \ne i_r=1}^\np
  \delta(\bmp_1 - \bmP_{i_1}^a)\,
  \cdots
  \delta(\bmp_r - \bmP_{i_r}^a)\,.
\end{align}
As in Eq.~(\ref{dtc}), the number of events in a subsubsample $\nevnpN
= \sum_a \delta(\np,\nsp(a))\, \delta(N , N(a))$ enters the relevant
event averages
\begin{align}
  \lleq{vdh}
  \rho(\bmp_1 ,\ldots,\bmp_r \cond \samplenpN)
  &= \frac{1}{\nevnpN} \sum_a
  \hro(\bmp_1 ,\ldots,\bmp_r \cond \np,N,\bmP^a)
\end{align}
where once again the double Kronecker deltas in (\ref{vdg}) ensure
selection of events in $\samplenpN$ only.  Integrals of the counters
over the good-azimuth region $\azia$ yield, for the eventwise and
$\samplenpN$-averaged counters,
\begin{align}
  \lleq{vdi}
  \int_\azia d\bmp_1 \cdots d\bmp_r\;
  \hro(\bmp_1 ,\ldots,\bmp_r \cond \np,N,\bmP^a)
  &= \np^{\ff{r}} \;\delta(\np,\nsp(a))\; \delta(N , N(a))\\
  \lleq{vdj}
  \int_\azia d\bmp_1 \cdots d\bmp_r\;
  \rho(\bmp_1 ,\ldots,\bmp_r \cond \samplenpN)
  &= \np^{\ff{r}}.
\end{align}
Bearing in mind that observation points $\bmp_1,\bmp_2,\ldots$ refer
to positive pions in $\azia$ only, the event-averaged counters for
fixed $N$ but any $\np$ are given by the average weighted in terms of
the relative frequency $\evrnpN = \nevnpN/\nevN$,
\begin{align}
  \lleq{vqd}
  \rho(\bmp_1,\ldots,\bmp_r\cond \sampleN) 
  &= \sum_{\np=r}^N \evrnpN\; \rho(\bmp_1,\ldots,\bmp_r\cond \samplenpN) 
  = \savN{\rho(\bmp_1,\ldots,\bmp_r\cond \samplenpN)}
\end{align}
for $r = 1,2,3,\ldots$, which integrate to
\begin{align}
  \lleq{vqf}
  \int_\azia  d\bmp_1\cdots d\bmp_r
  \;\rho(\bmp_1,\ldots,\bmp_r\cond \sampleN) 
  = \savN{n^{\ff{r}}}.
\end{align}

\section{Construction of correlation quantities}
\label{sec:corrq}

\subsection{Criteria}
\label{sec:ctra}

Correlation measurements of any sort are only meaningful if a
reference baseline signifying ``independence'' or ``lack of
correlation'' is defined quantitatively; indeed, many different kinds
of correlations may be defined and measured on the same data,
depending on which particular physical and mathematical scenario is
considered to be known or trivial and taken to be the
baseline \cite{2001-Eggers-Tihany-hepex0102005}. In our case, we
require the reference distribution to have the following properties:
\begin{enumerate}
\item \textit{The number of charged pions in all phase space $N$ is an
    important parameter as a measure of possibly different physics,
    but only the $n$ positive pions in $\azia$ are to be considered in
    the differential analysis.} 

\item \textit{For a given $(N,n)$, the momenta of the reference
    density $\rhoref(\bmp_1,\ldots,\bmp_r\cond \samplenpN)$ should be
    mutually independent for any order $1\leq r\leq n$.}  This and the
  previous requirement imply that the reference should be a
  $n$-multinomial distributed over continuous momentum space; see
  Section \ref{sec:muno}.

\item \textit{Given fixed $N$, the reference density
    $\rhoref(\bmp_1,\ldots,\bmp_r\cond \sampleN)$ must reproduce the
    $n$-multiplicity structure of the subsubsamples $\samplenpN$ as
    embodied in $\evrnpN$.} As set out further in Section
  \ref{sec:mnfn}, this translates into an \textit{average of
    multinomials},
  \begin{align}
    \lleq{vqg}
    \rhoref(\bmp_1,\ldots,\bmp_r\cond \sampleN)
    &= \sum_{\np=r}^N \evrnpN 
    \;\rhomult(\bmp_1,\ldots,\bmp_r\cond \bm{\alpha}, \samplenpN)
    = \savN{\rhomult(\bmp_1,\ldots,\bmp_r\cond \bm{\alpha}, \samplenpN)}.
  \end{align}

\item \textit{The reference density should reproduce the measured
    one-particle density in momentum space.} This can in principle be satisfied by
  three different expressions for the multinomial's parameters
  $\bmalpha$: see Section \ref{sec:mnalpha}.

\item \textit{Measures of correlation must reduce to zero even on a
    differential basis whenever the data is, in fact, uncorrelated.}
  While this may seem self-evident, this requirement is often ignored
  or not satisfied in the literature. We address the resulting proper
  baseline through the use of \textit{internal cumulants} in Section
  \ref{sec:intc}.

\item \textit{The measure of correlation should be insensitive to the
    one-particle distribution}.  This is addressed as usual by
  normalisation; see Section \ref{sec:intc}.
\end{enumerate}

\subsection{The reference distribution}

\subsubsection{Multinomials in discrete and continuous spaces}
\label{sec:muno}

Before Eq.~(\ref{vqg}) can be developed further, it is necessary to
take a detour into discrete outcome spaces before tackling the
continuous outcome space defined by $\bmp$ and $\bmP^a$. The reason is
that multinomial distributions for continuous arguments $\bmp$ can be
written only as a limit of the discrete precursor.

Let there be bins $\Omega_b, b=1,\ldots,B$ with the corresponding set
of \textit{Bernoulli probabilities} $\bmalpha = \{\alpha(b)\}_{b=1}^B$
of a single particle falling into bin $\Omega_b$, normalised by
$\sum_b \alpha(b) = 1$. Independent tossing of $n$ particles into
these bins results in the multinomial for the bin counts $\bmn =
\{n_b\}_{b=1}^B$,
\begin{align}
  \lleq{vqh} p(\bmn\cond \bmalpha,n) &= n!\prod_{b=1}^B
  \frac{\alpha(b)^{n_b}}{n_b!},
\end{align}
with normalisation
\begin{align}
  \lleq{vqha}
  \sum_{U(\bmn)} p(\bmn\cond \bmalpha,n) &= 1,
\end{align}
where the sum must be taken over the ``universal set''
\begin{align}
  \lleq{vqhb}
  U(\bmn) = \{\bmn \cond n_b \geq 0;\; \textstyle\sum_b n_b = n\}.
\end{align}
The multivariate factorial moment generating function (FMGF) for this
multinomial for the set of source parameters $\bmlambda =
\{\lambda(b)\}_{b=1}^B$ can be solved in closed form,
\begin{align}
  \lleq{vqi} %
  Q^{\rm mult}(\bmlambda\cond \bmalpha,n) 
  &= \sum_U p(\bmn\cond\bmalpha,n)\;\prod_b (1-\lambda(b))^{n_b}
  \ =\  \biggl[ 1 - \sum_b \lambda(b) \alpha(b) \biggr]^n.
\end{align}
The FMGF $Q(\bmlambda)$ can generally be used to find multivariate
factorial moments $\rho(b_{i_1},b_{i_2},\cdots,b_{i_r})$ and factorial
cumulants $\kappa(b_{i_1},b_{i_2},\cdots,b_{i_r})$ for any selection
of bins $(b_{i_1},b_{i_2},\cdots,b_{i_r}) \in (1,\ldots,B)$, including
repeated indices, by differentiation
\begin{align}
  \lleq{vqj}
  \rho(b_{i_1},b_{i_2},\cdots,b_{i_r}) &=  
  \frac{(-1)^r\; \partial^r Q(\bmlambda)}
  {\partial\lambda(b_{i_1})\;\partial\lambda(b_{i_2})\,\cdots\,\partial\lambda(b_{i_r})}
  \biggr|_{\bmlambda=0}, \\
  \lleq{vqk}
  \kappa(b_{i_1},b_{i_2},\cdots,b_{i_r}) &=  
  \frac{(-1)^r\; \partial^r\ln Q(\bmlambda)}
  {\partial\lambda(b_{i_1})\;\partial\lambda(b_{i_2})\,\cdots\,\partial\lambda(b_{i_r})}
  \biggr|_{\bmlambda=0}.
\end{align}
For the multinomial case (\ref{vqi}), the factorial moments and
cumulants are therefore
\begin{align}
  \lleq{vql}
  \rho^{\rm mult}(b_{i_1},b_{i_2},\cdots,b_{i_r}\cond \bmalpha,n) 
  &= n^{\ff{r}}\;\prod_{d=1}^r \alpha(i_d) \qquad \forall\;r\leq n,\\
  \lleq{vqm}
  \kappa^{\rm mult}(b_{i_1},b_{i_2},\cdots,b_{i_r}\cond \bmalpha,n) 
  &= (-1)^{r-1}\,(r-1)! \cdot n\prod_{d=1}^r \alpha(i_d).
\end{align}
The \textbf{multinomial for variable $\bmp$ in continuous outcome space}
$\mathbb{R}$ is derived by keeping $n$ constant while taking the
limit $B\to\infty$ with bin sizes tending to zero and changing to a
Bernoulli probability density $\alpha(b) \to d\bmp\,\alpha(\bmp)$
normalised by $\int_\azia d\bmp\,\alpha(\bmp) = 1$. The result is the
\textit{point process} where the probability for the count $n(\bmp)$
in the infinitesimal ``bin'' around any $\bmp$ to be larger than 1
becomes negligible, i.e.\ we have at most one particle at a given
$\bmp$. While the multinomial probability itself can be written only
as a limit, the FMGF can be written analytically as the functional
\cite{1972-KobaNielsenOlesen}
\begin{align}
  \lleq{vqn}
  Q^{\rm mult}[\lambda(\bmp)\cond \alpha(\bmp),n]
  = \left[ 1 - \int_\azia d\bmp\,\lambda(\bmp)\,\alpha(\bmp)\right]^n.
\end{align}
Factorial moments and factorial cumulants are found generically
from functional derivatives \cite{1973-Koba-appb4.95-xx.xx}
\begin{align}
  \lleq{vqo}
  \rho(\bmp_{i_1},\bmp_{i_2},\cdots,\bmp_{i_r}) &=  
  \frac{(-1)^r\; \delta^r Q[\lambda(\bmp)]}
  {\delta\lambda(\bmp_{i_1})\; \delta\lambda(\bmp_{i_2})\; 
    \cdots \delta\lambda(\bmp_{i_r})\;}
  \biggr|_{\lambda(\bmp)=0}, \\
  \lleq{vqp}
  \kappa(\bmp_{i_1},\bmp_{i_2},\cdots,\bmp_{i_r}) &=  
  \frac{(-1)^r\; \delta^r \ln Q[\lambda(\bmp)]}
  {\delta\lambda(\bmp_{i_1})\; \delta\lambda(\bmp_{i_2})\; 
    \cdots \delta\lambda(\bmp_{i_r})\;}
  \biggr|_{\lambda(\bmp)=0}, 
\end{align}
which for the multinomial $Q[\bmlambda(\bmp)] = Q^{\rm
  mult}[\lambda(\bmp)\cond \alpha(\bmp),n]$ of (\ref{vqn}) yield
\begin{align}
  \lleq{vqq}
  \rho^{\rm mult}(\bmp_{i_1},\bmp_{i_2},\cdots,\bmp_{i_r}\cond \alpha(\bmp),n) 
  &= n^{\ff{r}}\;\prod_{k=1}^r \alpha(\bmp_{i_k}) \qquad 1 \leq r\leq n,\\
  \lleq{vqr}
  \kappa^{\rm mult}(\bmp_{i_1},\bmp_{i_2},\cdots,\bmp_{i_r}\cond \alpha(\bmp),n) 
  &= (-1)^{r-1}\,(r-1)! \cdot n\prod_{k=1}^r \alpha(\bmp_{i_k}).
\end{align}

\subsubsection{Multinomial reference for fixed $N$}
\label{sec:mnfn}

Applying the above general case to our reference distribution
(\ref{vqg}), we must rewrite Eq.~(\ref{vqn}) to make provision for the
fact that $\bmalpha$ may in general depend not only on $N$ but also on
$n$,
\begin{align}
  \lleq{mnc}
  Q^{\rm mult}[\lambda(\bmp)\cond \alpha(\bmp\cond \samplenpN)]
  = \left[ 1 - \int d\bmp\,\lambda(\bmp)\,
    \alpha(\bmp\cond \samplenpN)\right]^{\np}.
\end{align}
Inserting (\ref{mnc}) into (\ref{vqg}), we find the FMGF for the
reference distribution of subsample $\sampleN$ to be
\begin{align}
  \lleq{mne}
  Q^{\rm ref}[\lambda(\bmp)\cond \alpha(\bmp),\sampleN ]
  &= \sum_\np \evrnpN \;
  Q^{\rm mult}[\lambda(\bmp)\cond \alpha(\bmp\cond \samplenpN)] %
  \ =\ \savNl{
  \left[ 1 - \int d\bmp\,\lambda(\bmp)\,
    \alpha(\bmp\cond \samplenpN)\right]^{\np} }.
\end{align}
Using (\ref{vqq}), the reference factorial moments are therefore,
\begin{align}
  \lleq{mnf}
  \rho^{\rm ref}(\bmp_1,\bmp_2,\cdots,\bmp_r\cond \sampleN) 
  &= \savNl{ \np^{\ff{r}}\;\prod_{k=1}^r \alpha(\bmp_k\cond \samplenpN) }
  \qquad \forall\;r\leq n
\end{align}
with corresponding expressions for the reference factorial cumulants.

\subsubsection{Reproducing the one-particle distribution}
\label{sec:mnalpha}

The set of functions $\alpha(\bmp\cond \samplenpN)$ are as yet
undetermined, apart from the general constraints $\alpha(\bmp\cond
\samplenpN) \geq 0$ and $\int_\azia d\bmp\,\alpha(\bmp\cond
\samplenpN)=1$. In multinomials of all kinds, the Bernoulli
probabilities $\bmalpha$ are fixed parameters and therefore are the
conveyers of whatever remains constant in the outcomes while the
detailed outcomes fluctuate as statistical outcomes do. The ``field''
$\alpha(\bmp\cond \samplenpN)$ can and must therefore be seen as the
quantity encompassing the ``physics'' of the one-particle
distributions, which, in the absence of additional external
information, is embodied by our experimental data sample: the
experimental densities $\rho(\bmp_1,\ldots,\bmp_\smallN\cond\sampleN)$ ``are''
the physics, including all correlations, and their first-order
projections $\rho(\bmp_1\cond\sampleN)$ ``are'' the one-particle
physics. The question immediately arises whether $\alpha(\bmp\cond
\samplenpN)$ should be fixed by $\rho(\bmp\cond \samplenpN)$ or the
$n$-average $\rho(\bmp\cond \sampleN) = \savN{\rho(\bmp\cond
  \samplenpN)}$.  Three possible choices come to mind:
\begin{enumerate}
\item It is tempting to define it in terms of the density for each
  subsubsample $\samplenpN$,
  \begin{align}
    \lleq{mnh}
    \alpha(\bmp\cond \samplenpN) &= \frac{\rho(\bmp\cond\samplenpN)}{n} 
    \qquad \forall\; (N,n),
  \end{align}
  which is correctly normalised since $\int d\bmp\,\rho(\bmp\cond \samplenpN) = n$.  As this
  choice would attribute physical significance to $n$, it would be
  appropriate whenever $n$ is associated with additionally measured
  experimental information. If, however, $n$ fluctuates randomly from
  event to event based in part on unmeasured or unmeasurable
  properties such as an event's azimuthal orientation, use of
  (\ref{mnh}) makes no sense.

  If $n$ is deemed physically relevant, correlations in terms of
  $\rho(\bmp_1 ,\ldots,\bmp_r \cond \samplenpN)$ of Eq.~(\ref{vdh})
  may be feasible, conditional on the availability of a sufficient
  number of events $\nevnpN$.  Where sample sizes do not permit this,
  one could nevertheless attempt to measure what have historically
  been termed ``short-range correlations'' but in this case not in the
  traditional sense of fixed-$N$ correlations versus inclusive ones,
  but rather of fixed-$n$-fixed-$N$ correlations versus
  fluctuating-$n$-fixed-$N$ correlations.  See Section \ref{sec:cnnp}.

\item A second choice
  \begin{align}
    \lleq{mni}
    \alpha(\bmp\cond \samplenpN) 
    &= \savNl{\frac{\rho(\bmp\cond\samplenpN)}{n}}
  \end{align}
  would be properly normalised but fails to satisfy the crucial
  relations (\ref{kqh})--(\ref{kqka}) below and is hence discarded.
  
\item While remaining open-minded towards Choice 1, we therefore
  choose the third possibility, the ratio of the average density
  divided by the average, all for fixed $N$,
  \begin{align}
    \lleq{mnj}
    \alpha(\bmp\cond \samplenpN) 
    &= \frac{\savN{\rho(\bmp\cond \samplenpN)}}{\savN{n}}
    = \frac{\rho(\bmp\cond \sampleN)}{\savN{n}}
  \end{align}
  which would be appropriate for samples where $\nevnpN$ is too small
  or physical significance can be attributed only to $N$ but not to
  $n$. According to (\ref{vqf}), it is also correctly normalised and
  ensures that the Bernoulli parameters are the same for all events in
  $\sampleN$, independent of $n$.  Substituting this into
  Eq.~(\ref{mnf}), the differential reference factorial moments orders
  become
  \begin{align}
    \lleq{mnk} %
    \rho^{\rm ref}(\bmp_1,\bmp_2,\cdots,\bmp_r\cond \sampleN)
    &= \frac{\savN{n^{\ff{r}}}}{\savN{n}^r} \;\prod_{d=1}^r \rho(\bmp_d\cond\sampleN)
    \ = \ F_{r\smallN} \;\prod_{d=1}^r \rho(\bmp_d\cond\sampleN)
  \end{align}
  where we identify the prefactor as the normalised factorial
  moments of the $n$-multiplicity distribution for given $N$,
  \begin{align}
    \lleq{mns}
    F_{r\smallN} = \frac{\savN{n^{\ff{r}}}} {\savN{n}^r},
  \end{align}
  while the generating functional (\ref{mne}) becomes (see also
  \cite{1999-SuzukiBiyajima-prc60.034903-hepph9907348})
  \begin{align}
    \lleq{mnea}
    Q^{\rm ref}[\lambda(\bmp)\cond \alpha(\bmp),\sampleN ]
    &= \savNl{ \left[ 1 - 
        \int d\bmp\,\lambda(\bmp)\,\frac{\rho(\bmp\cond \sampleN)}{\savN{n}}\right]^{\np} }.
  \end{align}
\end{enumerate}
Taking functional derivatives of the logarithm of (\ref{mnea}), the
first, second and third order cumulants of the reference density are
\begin{align}
  \lleq{mnl}
  \kappa^{\rm ref}(\bmp_1\cond \sampleN) 
  &= \rhoref(\bmp_1\cond\sampleN)\ =\ \rho(\bmp_1\cond\sampleN) \\
  \lleq{mnm} 
  \kappa^{\rm ref}(\bmp_1,\bmp_2\cond \sampleN) 
  &= \rhoref(\bmp_1,\bmp_2\cond \sampleN)
  - \rho(\bmp_1\cond \sampleN)\; \rho(\bmp_2\cond \sampleN)  \\
  \lleq{mnn}
  &= \left(\frac{\savN{\np^\ff{2}}}{\savN{\np}^2} - 1 \right)
  \rho(\bmp_1\cond \sampleN)\,\rho(\bmp_2\cond \sampleN) 
  \\
  \lleq{mno}
  \kappa^{\rm ref}(\bmp_1,\bmp_2,\bmp_3\cond \sampleN) 
  &= \rhoref(\bmp_1,\bmp_2,\bmp_3\cond \sampleN)
  \\
  &\quad - [3]\,\rhoref(\bmp_1,\bmp_2\cond \sampleN) 
  \rho(\bmp_3\cond \sampleN)
  +2 
  \rho(\bmp_1\cond \sampleN)
  \rho(\bmp_2\cond \sampleN)
  \rho(\bmp_3\cond \sampleN)
  \nonumber\\
  \lleq{mnp}
  &= \left(
    \frac{\savN{\np^{\ff{3}}}} {\savN{\np}^3}\;
    - 3 \frac{\savN{\np^{\ff{2}}}} {\savN{\np}^2}
    + 2
    \right)
   \rho(\bmp_1\cond \sampleN) 
   \rho(\bmp_2\cond \sampleN) 
   \rho(\bmp_3\cond \sampleN) 
\end{align}
where the square bracket $[3]$ indicates the number of distinct
permutations which must be taken into account.  The terms in the
rounded brackets are readily recognised as the normalised factorial
cumulants of the $\np$ distribution for a given fixed $N$
\begin{align}
  \lleq{mnq}
  K_{r\smallN} &=
  \frac{1}{\savN{n}^r}
  \frac{(-\partial)^r}{\partial\Lambda^r}\ln\left(
    \sum_{\np=0}^N \evrnpN\,(1-\Lambda)^n
    \right)_{\Lambda=0},
\end{align}
and so generalisation to arbitrary orders is immediate,
\begin{align}
  \lleq{mnr}
  \kappa^{\rm ref}(\bmp_1,\cdots,\bmp_r\cond \sampleN) 
  = K_{r\smallN} \prod_{k=1}^r \rho(\bmp_k\cond \sampleN).
\end{align}
This can be proven generally by defining the functional
$\Lambda[\lambda(\bmp)] = \int d\bmp\,\lambda(\bmp)\,\rho(\bmp\cond
\sampleN)/ \savN{n}$ which has only a first nonzero functional
derivative $\delta\Lambda/\delta\lambda(\bmp_1)
=\rho(\bmp\cond\sampleN)/\savN{n}$ and the multiplicity generating
function ${\mathcal Z}(\Lambda\cond \sampleN) = \sum_n \evrnpN
(1-\Lambda)^n$, in terms of which $Q^{\rm ref}[\lambda] = {\mathcal
  Z}[\Lambda[\lambda]]$.

\subsection{Internal cumulants for fixed $\sampleN$}
\label{sec:intc}

Eq.~(\ref{mnr}) shows that the differential cumulants of the reference
distribution are directly proportional to the integrated cumulants
$K_{r\smallN}$ of $n$, which are zero only if $\evrnpN$ is
poissonian. For fixed $N$, neither the integrated cumulants $K_{r\smallN}$ 
nor the differential ones are zero. While this has long been
recognised in the literature \cite{1975-Foa-pr22.1-xx.xx}, the
inevitable consequence was not drawn, namely that ``poissonian''
cumulants for fixed $N$
\begin{align}
  \lleq{kqd}
  \kappa(\bmp_1,\bmp_2\cond \sampleN) 
  &= \rho(\bmp_1,\bmp_2\cond \sampleN) -
  \rho(\bmp_1\cond \sampleN)\; \rho(\bmp_2\cond \sampleN), \\
  \lleq{kqe}
  \kappa(\bmp_1,\bmp_2,\bmp_3\cond \sampleN) 
  &= \rho(\bmp_1,\bmp_2,\bmp_3\cond \sampleN) \nonumber\\
  &\quad - [3]\, 
  \rho(\bmp_1,\bmp_2\cond \sampleN) \; \rho(\bmp_3\cond \sampleN) 
  + 2   
  \rho(\bmp_1\cond \sampleN)\;
  \rho(\bmp_2\cond \sampleN)\;
  \rho(\bmp_3\cond \sampleN)
\end{align}
etc.\ cannot possibly represent true correlations because they are
nonzero even when the momenta are fully independent.  It is known that
the theory of cumulants needs improvement on a fundamental level which
reaches well beyond the scope of this paper \cite{Ken86a},
\cite{2009-DeKock-MSc}, but those difficulties are irrelevant here. A
first step which does address the above concerns was taken in
Ref.~\cite{1996-LipaEggersBuschbeck-prd53.4711-hepph9604373}, where it
was shown very generally on the basis of generating functionals that
correlations for samples of fixed $N$ are best measured using the
\textit{internal cumulants} $\kappa^I$, which are defined as the
differences between the measured and the reference cumulants of the
same order
\begin{align}
  \lleq{kqea} \kappa^I(\bmp_1,\ldots,\bmp_r\cond\sampleN) %
  &= \kappa(\bmp_1,\ldots,\bmp_r\cond\sampleN) -
  \kappa^{\rm ref}(\bmp_1,\ldots,\bmp_r\cond\sampleN).
\end{align}
For our averaged-multinomial reference case, the internal cumulants of
second and third order are given by the differences between
Eqs.~(\ref{kqd}) and (\ref{mnn}) and between (\ref{kqe}) and
(\ref{mnp}), resulting in
\begin{align}
  \lleq{kqf}
  \kappa^I(\bmp_1,\bmp_2\cond \sampleN) 
  &= \rho(\bmp_1,\bmp_2\cond \sampleN)
  - F_{2\smallN}
    \rho(\bmp_1\cond \sampleN)\;\rho(\bmp_2\cond \sampleN),
  \\
  \lleq{kqg}
  \kappa^I(\bmp_1,\bmp_2,\bmp_3\cond \sampleN) 
  &= \rho(\bmp_1,\bmp_2,\bmp_3\cond \sampleN) 
  - [3]\,\rho(\bmp_1,\bmp_2\cond \sampleN) \; \rho(\bmp_3\cond \sampleN) 
  \nonumber\\
  &\quad + G_{3\smallN} \; \rho(\bmp_1\cond \sampleN) \rho(\bmp_2\cond \sampleN) 
   \rho(\bmp_3\cond \sampleN),
\end{align}
with
\begin{align}
  \lleq{kqgb}
  G_{3\smallN} &= 3 F_{2\smallN} - F_{3\smallN}
  \ =\ 3\,\frac{\savN{n(n{-}1)}}{\savN{n}^2} - \frac{\savN{n(n{-}1)(n{-}2)}}{\savN{n}^3}
\end{align}
and so on for higher orders.  These internal cumulants are identically
zero if and when the measured densities for fixed $\samplenpN$ are
multinomials since then from Eq.~(\ref{mnk})
\begin{align}
  \lleq{kqh}
  \rho(\bmp_1,\bmp_2\cond \sampleN)
  &\rightarrow \rhoref(\bmp_1,\bmp_2\cond \sampleN)
  = F_{2\smallN}
    \rho(\bmp_1\cond \sampleN)\;\rho(\bmp_2\cond \sampleN)
\end{align}
so that $\kappa^I(\bmp_1,\bmp_2\cond \sampleN) \rightarrow 0$ whenever
the data is multinomial, while
\begin{align}  
  \lleq{kqi}
  \rho(\bmp_1,\bmp_2,\bmp_3\cond \sampleN)
  \rightarrow
   F_{3\smallN}
   \rho(\bmp_1\cond \sampleN) 
   \rho(\bmp_2\cond \sampleN) 
   \rho(\bmp_3\cond \sampleN) 
\end{align}
ensures that $\kappa^I(\bmp_1,\bmp_2,\bmp_3\cond \sampleN) \rightarrow
0$ in the same case.  On another level, the internal cumulants always
integrate to zero over the full good-azimuth space $\azia$,
irrespective of the presence of correlations,
\begin{align}
  \lleq{kqka}
  \int_\azia d\bmp_1\,d\bmp_2\; \kappa^I(\bmp_1,\bmp_2\cond \sampleN) 
  \ =\
  \int_\azia d\bmp_1\,d\bmp_2\,d\bmp_3\; 
  \kappa^I(\bmp_1,\bmp_2,\bmp_3\cond \sampleN) 
  \ =\ 0
\end{align}
and so on for all orders. Both properties will remain valid after
transformation from three-momenta to invariant four-momentum
differences in Section \ref{sec:qrs}.  In the case of Poissonian
statistics, $F_{r\smallN} = 1\;\forall r$, so that the above internal
cumulants revert to their usual definitions.

As stated in Section \ref{sec:ctra}, the measured correlations may in
addition be made insensitive to the one-particle distribution through
normalisation.  As set out in
Ref.~\cite{1996-LipaEggersBuschbeck-prd53.4711-hepph9604373}, such
normalisation is achieved for fixed $N$ by dividing the internal
cumulants by the corresponding reference distribution density, which
for the case at hand is given by Eq.~(\ref{mnk}). This leads to the
second-order normalised internal cumulant
\begin{align}
  \lleq{ncd}
  K^I(\bmp_1,\bmp_2\cond \sampleN) 
  &= \frac{1}{F_{2\smallN}} \;
  \frac{\rho(\bmp_1,\bmp_2\cond \sampleN)}
    {\rho(\bmp_1\cond \sampleN)\;\rho(\bmp_2\cond \sampleN)}
    \ - \ 1,
\end{align}
while in third order we get
\begin{align}
  \lleq{nce}
  K^I(\bmp_1,\bmp_2,\bmp_3\cond \sampleN)
  &= \frac{1}{F_{3\smallN}}
  \left(
      \frac{
        \rho(\bmp_1,\bmp_2,\bmp_3\cond \sampleN) 
        - [3]\,\rho(\bmp_1,\bmp_2\cond \sampleN) 
        \; \rho(\bmp_3\cond \sampleN) 
      }
      { \rho(\bmp_1\cond \sampleN) 
        \rho(\bmp_2\cond \sampleN) 
        \rho(\bmp_3\cond \sampleN) 
      } 
    \right)
    + 3 \frac{F_{2\smallN}}{F_{3\smallN}}
      - 1.
\end{align}

\subsection{Correlation integrals for momentum differences}
\label{sec:qrs}

In femtoscopy, correlations are mostly expressed in terms of pair
variables $\bmK = \tfrac{1}{2}\left( \bmp_1 + \bmp_2\right)$ and
difference $\bmq = \bmp_1 - \bmp_2$ or the invariant four-momentum
\cite{1995-AndreevBiyajimaDreminSuzuki-ijmpa10.3951-hepph9501345} %
$Q = %
\sqrt{-(p_1 - p_2)^2} = \sqrt{(\bmp_1-\bmp_2)^2 - (E_1-E_2)^2}$ where
the energies are on-shell, $E_r = \sqrt{\bmp_r^2+m^2}$. As shown in
Ref.~\cite{1993-EggersLipaCarruthersBuschbeck-plb301.298-xx.xx}, the
formulation of event\-wise counters as sums and products of Dirac
delta functions makes it easy to change variables.  Writing
$\rhon{r}(\bmp_1,\ldots,\bmp_r)$ as shorthand for
$\rho(\bmp_1,\ldots,\bmp_r\cond \sampleN)$ etc., the second-order
unnormalised internal cumulant in terms of $Q$ is, for example, found
from the identity $\int dQ\,\kappan{2}^I(Q) = \int dQ\,\int_\azia
d\bmp_1\,d\bmp_2\,\kappan{2}^I(\bmp_1,\bmp_2)$ $\delta(Q -
\sqrt{(\bmp_1-\bmp_2)^2 - (E_1-E_2)^2})$ to be
\begin{align}
  \lleq{trk}
  \kappan{2}^I(Q) 
  &= \savNal{ \sum_{i\ne j} \delta(Q - Q_{ij}^{aa})}
  - F_{2\smallN} 
  \savNal{\savNb{\sum_{i,j} \delta(Q - Q_{ij}^{ab})}} 
  \\
  \lleq{trl}
  &= \rhon{2}(Q) - 
  F_{2\smallN}\, \rhon{1}{\otimes}\rhon{1}(Q),
\end{align}
where the counters in the second line are defined by the terms in the
first, while $Q_{ij}^{aa} = [-(P_i^a-P_j^a)^2]^{1/2}$ and $Q_{ij}^{ab} =
[-(P_i^a-P_j^b)^2]^{1/2}$ are four-momentum differences between
sibling pairs $aa$ and event mixing pairs $ab$ respectively.  It is
easy to show that $\int_0^\infty dQ\,\rhon{2}(Q) =
\savN{\np^{\ff{2}}}$ and $\int_0^\infty
dQ\,\rhon{1}{\otimes}\rhon{1}(Q) = \savN{\np}^2 $ and hence, as
before, $\int_0^\infty dQ\,\kappan{2}^I(Q) =0$, which will be true for
any correlation whatsoever.  The double event averages in the product
term
\begin{align}
  \lleq{tro}
  \rhon{1}{\otimes}\rhon{1}(Q)
  = \savNal{\savNb{\sum_{i,j} \delta(Q - Q_{ij}^{ab})}} 
\end{align}
are the theoretical foundations of event mixing
\cite{1993-EggersLipaCarruthersBuschbeck-plb301.298-xx.xx}; the inner
$b$-average is usually shortened to a smaller ``moving average tail''
subsample of $\sampleN$.

In third order, the ``GHP average'' invariant is defined as the
average of three two-momentum differences over all pairs (with or 
without the $\sqrt{3}$),
\begin{align}
  \lleq{trp}
  \qghpa = \sqrt{-(p_1-p_2)^2 - (p_2-p_3)^2 - (p_3-p_1)^2}\;/\sqrt{3}\,;
\end{align}
it is related to the the invariant mass of three pions $M_3 =
(p_1+p_2+p_3)^2$ by $\qghpa^2 = \tfrac{1}{3}M_3^2-m^2$. Other
``topologies'' such as the ``GHP max'' $ \qghpm =
\sqrt{\max[-(p_1-p_2)^2 , -(p_2-p_3)^2 , -(p_3-p_1)^2]} $ can also be
employed. For large multiplicities, the ``Star'' topology may be
preferred
\cite{1993-EggersLipaCarruthersBuschbeck-prd48.2040-hepph9304208}, but
we shall not pursue it here.  For the GHP average, the third internal
cumulant is given by
\begin{align}
  \lleq{trs}
  \kappan{3}^I(\qghpa)
  &= \savNal{\sum_{i\ne j\ne k} \delta\left(\qghpa - Q_{ikj}^{aaa}\right)}
  - 3 \savNal{\savNbl{\sum_{i\ne j}\sum_k \delta\left(\qghpa - Q_{ijk}^{aab}\right)}}
  \\
  &\quad + G_{3\smallN} 
  \savNal{\savNbl{\savNcl{\sum_{i,j,k} \delta\left(\qghpa - Q_{ijk}^{abc}\right)}}},
  \nonumber
\end{align}
with $Q_{ijk}^{abc} = 
\sqrt{\tfrac{1}{3}[-(P_i^a - P_j^b)^2 - (P_j^b - P_k^c)^2 - (P_k^c -
  P_i^a)^2] }$, and similarly for $Q_{ikj}^{aaa}$ and $Q_{ikj}^{aab}$.
Second and third-order cumulants are normalised by, respectively,
\begin{align}
  \lleq{trt}
  F_{2\smallN}\;\rhon{1}{\otimes}\rhon{1}(Q)
  &=  F_{2\smallN} 
  \savNal{\savNbl{\sum_{i,j} \delta(Q - Q_{ij}^{ab})}},
  \\
  \lleq{tru}
  F_{3\smallN}\;\rhon{1}{\otimes}\rhon{1}{\otimes}\rhon{1}(\qghpa)
  &= F_{3\smallN} 
  \savNal{\savNbl{\savNcl{\sum_{i,j,k} \delta\left(\qghpa - Q_{ijk}^{abc}\right)}}}.
\end{align}
After transforming from momenta to $Q$, the formulae of Section
\ref{sec:intc} become
\begin{align}
  \lleq{tsc}
  \kappan{2}^I(Q)
  &= \rhon{2}(Q) - F_{2\smallN} \rhon{1}{\otimes}\rhon{1}(Q),
  \\
  \lleq{tsd}
  \kappan{3}^I(\qghpa)
  &= \rhon{3}(\qghpa) - [3] \rhon{2}{\otimes}\rhon{1}(\qghpa)
  + G_{3\smallN}\, \rhon{1}{\otimes}\rhon{1}{\otimes}\rhon{1}(\qghpa),
\end{align}
while the normalised cumulants of Section \ref{sec:intc} become
\begin{align}
  \lleq{tse}
  K_2^I(Q\cond\sampleN) 
  &= \frac{\rhon{2}(Q)}{F_{2\smallN} \rhon{1}{\otimes}\rhon{1}(Q)}
  \ -\ 1,
  \\
  \lleq{tsf}
  K_3^I(\qghpa\cond\sampleN) 
  &= \frac{\rhon{3}(\qghpa) - [3]\rhon{2}{\otimes}\rhon{1}(\qghpa)}
  {F_{3\smallN}\,\rhon{1}{\otimes}\rhon{1}{\otimes}\rhon{1}(\qghpa)}
    \ +\ 3 \frac{F_{2\smallN}}{F_{3\smallN}} - 1.
\end{align}

\subsection{Effect of fixed-$N$ correction factors}

To get a feeling for the size of the corrections involved, we measured
the correction factors $F_{r\smallN}$ and $G_{3\smallN}$ with the same
UA1 dataset and the same cuts as in Fig.~1. As shown in Fig.~2, the
consequence of the clearly subpoissonian multiplicity distributions
shown in Fig.~1 is that these factors are significantly \textit{less}
than 1, in contrast to the usual factorial moments of the charged
multiplicity distribution which are superpoissonian with factors
\textit{exceeding} 1. For very low multiplicities $N < 10$, normalised
internal cumulants are hence larger than their poissonian counterparts
but converge to them with increasing $N$. Nevertheless up to $N \simeq
30$ corrections of more than 5\% for $K_2^I$ and more than 20\% for
$K_3^I$ compared to their uncorrected counterparts can be expected. By
contrast, the additive correction $G_{3\smallN}$ does not deviate much
from the poissonian limit of 2 except for very small $N$. By contrast,
unnormalised internal cumulants (\ref{tsc})--(\ref{tsd}) are far less
sensitive to the multinomial correction.
\begin{figure}[htbp]
  \includegraphics[width=88mm]{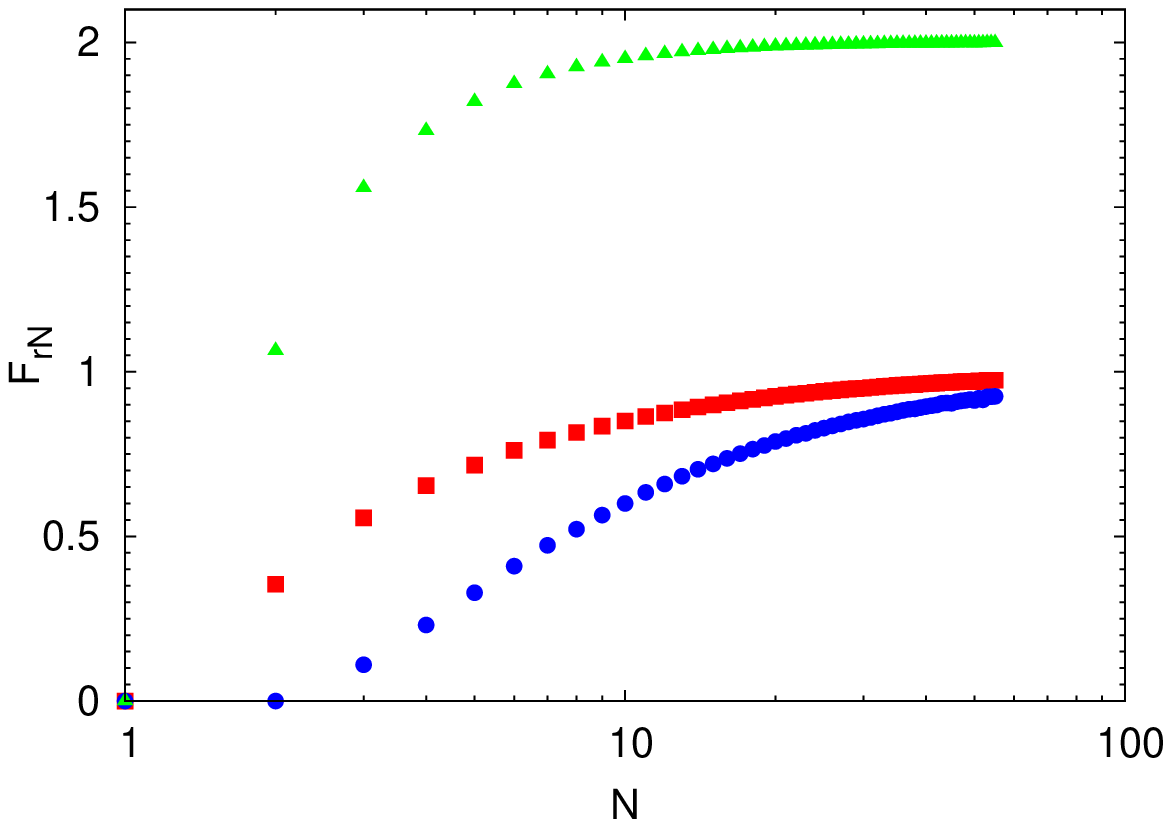}
  \hspace*{-12pt}
  \includegraphics[width=88mm]{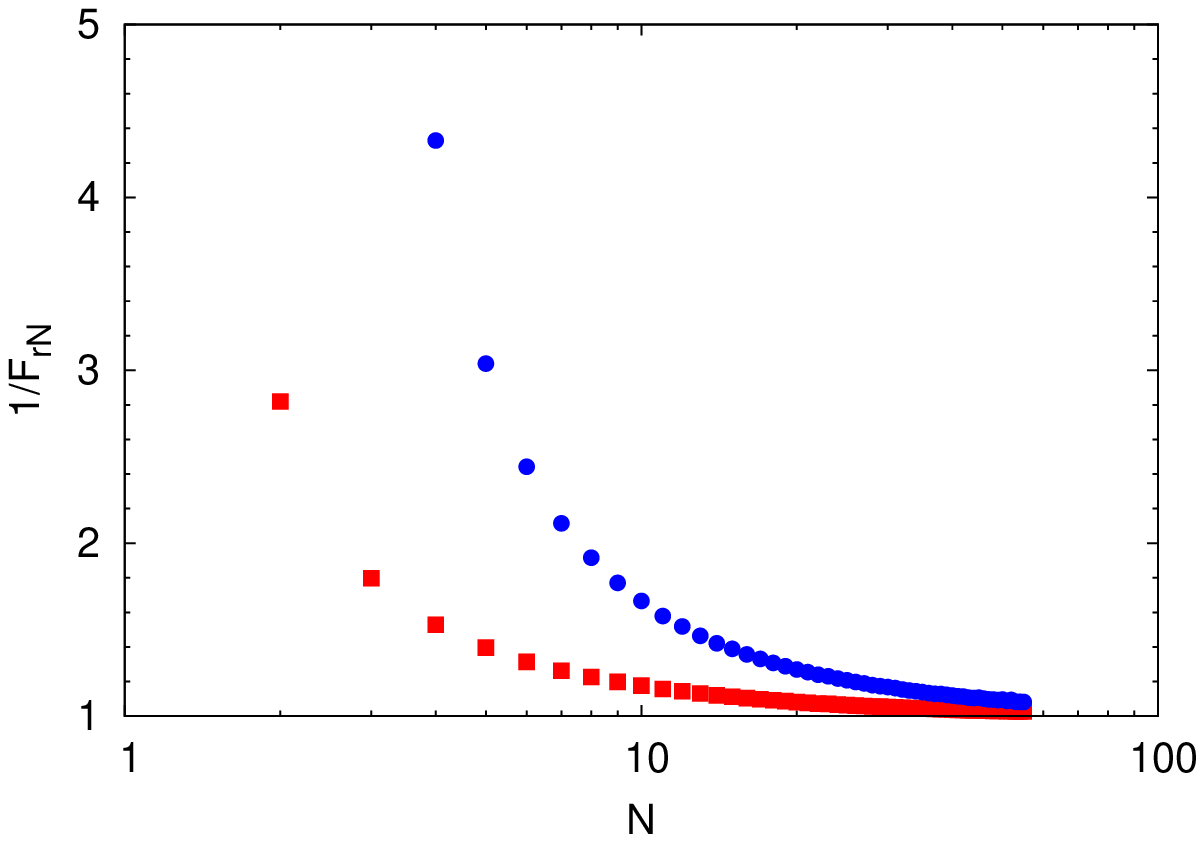}
  \caption{\small Left panel: Correction factors %
    $F_{2\smallN} = \savN{n(n{-}1)}/\savN{n}^2$ (red squares) and
    $F_{3\smallN} = \savN{n(n{-}1)(n{-}2)}/\savN{n}^3$ (blue circles)
    as defined in Eq.~(\ref{mns}) as well as $G_{3\smallN}$ (green
    triangles) of Eq.~(\ref{kqgb}), for the UA1 dataset used in
    Ref.~\cite{2006-EggersOctoberBuschbeck-plb635.280-hepex0601039}. Right
    panel: inverse factors as used in the normalisation of internal
    cumulants Eqs.~(\ref{ncd})--(\ref{nce}) and
    (\ref{tse})--(\ref{tsf}).}
\end{figure}

It is of interest to zoom in on the approach to the poissonian limit
of 1 and to compare these corrections to the equivalent
charged-multiplicity-based ones, which for the case of fixed $N$,
would be just $N(N{-}1)/N^2$ and $N(N{-}1)(N{-}2)/N^3$.  In Fig.~3,
the poissonian limit corresponds to zero on the $y$-axis.  It is clear
that the fixed-$N$ factors go some way to correct for the fixed-$N$
conditioning; the gap between them is approximately determined by
$\savN{n^{\ff{r}}}/N^{\ff{r}}$, i.e.\ by the exact definition and
outcome space for $n$.
\begin{figure}
  \centerline{\includegraphics[width=88mm]{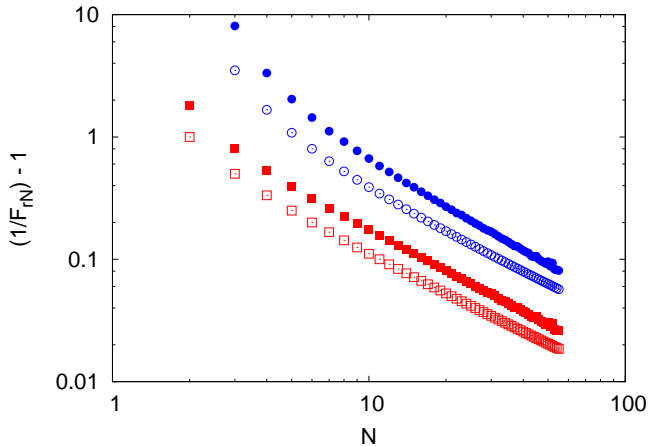}}
  \caption{\small Comparison, for the same UA1 data and cuts, of
    $(1/F_{2\smallN}) -1$ (red filled squares), $(1/F_{3\smallN}) -1$
    (blue filled circles), both averages over $n$ for fixed $N$, with
    prefactors $[N^2/N(N{-1})] - 1$ (red open squares) and
    $[N^3/N(N{-1})(N{-}2)] - 1$ (blue open circles). The poissonian
    limit is represented by zero on the $y$-axis, i.e.\ 100 times the
    $y$-scale represents the percentage deviation from the poissonian
    limit.}
\end{figure}

\subsection{Eliminating fluctuations in $\np$}
\label{sec:cnnp}

We return briefly to the first choice in Section \ref{sec:mnalpha}
i.e.\ $\alpha(\bmp\cond \samplenpN) = \rho(\bmp\cond\samplenpN)/n$
which would permit different physics for each $(N,\np)$
combination. If we were willing and able to do analyses for each
$\samplenpN$, we would use the fixed-$(N,n)$ equivalent of
(\ref{kqf})--(\ref{kqg}) derived from $ \kappa^{\rm
  ref}(\bmp_1,\ldots,\bmp_r \cond \samplenpN) = (-1)^{r-1}\,(r-1)!
\;\np \prod_{k=1}^r \rho(\bmp_k\cond\samplenpN)$,
\begin{align}
  \lleq{kqfn}
  \kappa^I(\bmp_1,\bmp_2\cond \samplenpN) 
  &= \rho(\bmp_1,\bmp_2\cond \samplenpN) - \left(1-\tfrac{1}{n}\right)
  \rho(\bmp_1\cond \samplenpN)\; \rho(\bmp_2\cond \samplenpN) \\
  \lleq{kqgn}
  \kappa^I(\bmp_1,\bmp_2,\bmp_3\cond \samplenpN) 
  &= \rho(\bmp_1,\bmp_2,\bmp_3\cond \samplenpN) 
  - [3]\, \rho(\bmp_1,\bmp_2\cond \samplenpN) \; \rho(\bmp_3\cond \samplenpN) 
  \nonumber\\
  &\quad + 2\left(1 - \tfrac{1}{n^2}\right)
  \rho(\bmp_1\cond \samplenpN)\;
  \rho(\bmp_2\cond \samplenpN)\;
  \rho(\bmp_3\cond \samplenpN)
\end{align}
and normalise by $\rho^{\rm ref}(\bmp_1,\ldots,\bmp_r\cond \samplenpN)
= (n^{\ff{r}}/n^r) \,\prod_{k=1}^r \rho(\bmp_k\cond\samplenpN)$.
Where that is not possible, we can still average over the above to
form ``Averaged Internal'' (AI) correlations (see Section
\ref{sec:smpab}),
but in this case averaging over $n$ for fixed $N$,
\begin{align}
  \lleq{kqfs}
  \kappa^{\rm\scriptscriptstyle SRC}(\bmp_1,\bmp_2\cond \sampleN) 
  &= \savN{\kappa^I(\bmp_1,\bmp_2\cond \samplenpN)}\\
  \lleq{kqgs}
  \kappa^{\rm\scriptscriptstyle SRC}(\bmp_1,\bmp_2,\bmp_3\cond \sampleN) 
  &= \savN{\kappa^I(\bmp_1,\bmp_2,\bmp_3\cond \samplenpN)}
\end{align}
and normalise if necessary by the moment $\rho^{\rm\scriptscriptstyle
  SRC}(\bmp_1,\ldots,\bmp_r\cond \sampleN) = \savN{(n^{\ff{r}}/n^r)
  \,\prod_{k=1}^r \rho(\bmp_k\cond\samplenpN)}$.  Given that this
involves products of moments in the subsubsample $\samplenpN$, event
mixing would have to be restricted to the same subsubsamples also; for
example
\begin{align}
  \savN{\left(1-\tfrac{1}{\np}\right)
  \rho(\bmp_1\cond \samplenpN)\, \rho(\bmp_2\cond \samplenpN) }
  &=
  \sum_{\np=2}^N \frac{\nevnpN}{\nevN}
  \left(1-\tfrac{1}{\np}\right)
    \frac{1}{(\nevnpN)^2}\sum_{a,b \,\in\, \samplenpN}
      \hro(\bmp_1\cond \np,N,\bmP^a)\,
      \hro(\bmp_2\cond \np,N,\bmP^b)
\end{align}
The transformation to pair variables works in the same way as in
previous sections.

\section{Statistical errors}
\label{sec:stderrors}

While the various versions of internal cumulants, constructed above
may all be relevant at some point, we concentrate on finding
expressions for experimental standard errors for the unnormalised and
normalised internal cumulants of Eqs.~(\ref{tsc})--(\ref{tsf}).  This
turns out to be more subtle than merely applying a generic
root-mean-square prescription.  We shall show in this section that
standard errors implemented thus far may have been underestimated even
in the standard two-particle case.

The calculations performed in this section belong to the
``frequentist'' view of probability; a proper Bayesian analysis, which
can be expected to rest on more solid foundations, is beyond the scope
of this paper. The two viewpoints can reasonably be expected to yield
similar results in the limit of large bin contents and sample sizes.

In this section, we often simplify notation by writing
$\rho_r(Q\cond\sampleN) \to \rhon{r}$ and
$\rho_1{\otimes}\rho_1(Q\cond\sampleN) \to \rhon{1}^2$ etc, since the
formulae apply to samples and variables of any kind.

\subsection{Propagation of statistical errors}

Because cumulants can be measured only through the moments that enter
their definitions, the first task is to identify which moment
variances and covariances are needed. By means of standard error
propagation \cite{Ken86a}, we find the sample variances for
second-order internal cumulants of Eqs.~(\ref{tsc}) and
Eq.~(\ref{tse}) to be
\begin{align}
  \lleq{ssc}
  \var(\kappa_2^I(Q\cond\sampleN))
    &= \var(\rhon{2} - F_{2\smallN}\, \rhon{1}^2)
    = \var(\rhon{2}) + F_{2\smallN}^2\,\var(\rhon{1}^2) - 2\,F_{2\smallN}\,\cov(\rhon{2},\rhon{1}^2) \\
    \lleq{ssd}
    \var(K_2^I(Q\cond \sampleN) )
    &= \var\left( \frac{1}{F_{2\smallN}} \frac{\rhon{2}} {\rhon{1}^2} -1  \right) \\
    \lleq{ssda}
    &= \left[ \frac{1}{F_{2\smallN}} \frac{\rhon{2}} {\rhon{1}^2} \right]^2
    \left[ %
         \frac{\var(\rhon{2})} {\rhon{2}^2}
      +  \frac{\var(\rhon{1}^2)} {\rhon{1}^4}
      - 2\frac{\cov(\rhon{2},\rhon{1}^2)} {\rhon{2}\cdot\rhon{1}^2}
    \right] 
\end{align}
under the assumption that $\var(F_{2\smallN})$ is much smaller than
the other variances, so that $F_{2\smallN}$ can be treated as a
constant; this is the case if there are many bins for $Q$.  Similarly,
from (\ref{kqg}), the variance of the unnormalised internal cumulant
is, assuming $G_3$ of Eq.~(\ref{kqgb}) to be constant,
\begin{align}
  \lleq{ssf}
  \var(\kappa_3^I(\qghpa\,|\,\sampleN)) 
  &= \var\left[\rhon{3} - 3\,\rhon{2}\rhon{1} + G_3\,\rhon{1}^3\right]
  \\
  &= \var(\rhon{3}) + 9\,\var(\rhon{2}\rhon{1}) + G_3^2 \var(\rhon{1}^3)
  \nonumber\\
  \lleq{ssg}
  &\quad - 6\,\cov(\rhon{3},\rhon{2}\rhon{1}) + 2G_3\,\cov(\rhon{3},\rhon{1}^3)
  - 6G_3\,\cov(\rhon{2}\rhon{1},\rhon{1}^3),
\end{align}
while the normalised version has variance (again assuming $\var(G_3) \ll
\var\rho_r$)
\begin{align}
  \lleq{ssh}
  \var(K_3^I(\qghpa\,|\,\sampleN))
  &= \var\left[ \frac{\rhon{3}-3\rhon{2}\rhon{1} + G_3\rhon{1}^3}{F_{3\smallN}\rhon{1}^3} \right]
  = \frac{1}{F_{3\smallN}^2} 
  \var\left[ \frac{\rhon{3}-3\rhon{2}\rhon{1}}{\rhon{1}^3} + G_3 \right]
  \\
  &= \left[\frac{\rhon{3}-3\rhon{2}\rhon{1}}{F_{3\smallN}\,\rhon{1}^3}\right]^2
  \left[ \frac{\var(\rhon{3}-3\rhon{2}\rhon{1})}{(\rhon{3}-3\rhon{2}\rhon{1})^2}
    + \frac{\var(\rhon{1}^3)}{\rhon{1}^6}
    - \frac{2\cov(\rhon{3}-3\rhon{2}\rhon{1},\rhon{1}^3)}{(\rhon{3}-3\rhon{2}\rhon{1})\rhon{1}^3}
  \right] 
  \nonumber\\
  \lleq{ssi}
  &= \left[\frac{\rhon{3}-3\rhon{2}\rhon{1}}{F_{3\smallN}\,\rhon{1}^3}\right]^2 
  \biggl[
    \frac{\var(\rhon{3})+9\var(\rhon{2}\rhon{1})-6\cov(\rhon{3},\rhon{2}\rhon{1})}
    {(\rhon{3}-3\rhon{2}\rhon{1})^2}
    + \frac{\var(\rhon{1}^3)}   {\rhon{1}^6}
  \nonumber\\ 
  &\hspace*{96pt}
    + \frac{6\cov(\rhon{2}\rhon{1},\rhon{1}^3) - 2\cov(\rhon{3},\rhon{1}^3) }
    {(\rhon{3}-3\rhon{2}\rhon{1})\rhon{1}^3}
    \biggr].
\end{align}
The unnormalised cumulants (\ref{tsc})--(\ref{tsd}) and their
variances (\ref{ssc}) and (\ref{ssg}) require knowledge of the
multiplicity factorial moments $F_{2\smallN}, F_{3\smallN}$, so that the
individual terms must be accumulated until the entire sample has been
analysed. By contrast, the normalised cumulants
(\ref{tse})--(\ref{tsf}) and their variances (\ref{ssda}) and
(\ref{ssi}) contain the multiplicity moments only as prefactors.

\subsection{Expectation values of counters}
\label{sec:exct}

While we shall not make direct use of the results in this section, it
is nevertheless useful briefly to consider what we might mean by an
``expectation value of experimental counters and densities''.  For any
scalar function $f(\bmp)$ of the momenta, the theoretical expectation
value $E[f]$ is defined as the integral over the entire outcome space
$\Omega$ of $f$ weighted by a ``parent distribution'' $P(\bmp)$, an
abstract entity supposedly containing everything there is to know on
this level,
\begin{align}
  \lleq{vrd}
  E[f(\bmp)] = \int_\Omega d\bmp\,P(\bmp)\,f(\bmp).
\end{align}
Purely theoretical concepts such as $P(\bmp)$ and $E[f]$ should be
given little or no room in a strongly experimentally-oriented
study. In calculating standard errors on counters below, we shall,
however, make use of the exact factorisation that expectation values
provide whenever two variables $x,y$ are statistically independent,
$E[xy] = \int dx\,dy\,P(x,y)\,xy = E[x]\,E[y]$.

Expectation values for pairwise variables such as the four-momentum
difference $Q$ we are considering here must be based on the underlying
physics. We can deduce some properties of the parent distribution
based on the usual definition of the femtoscopic correlation function
\begin{align}
  \lleq{vre} %
  C_2(Q) %
  &\equiv K_2(Q) + 1 = \frac{\rho_2^{\rm sibling}(Q)}{\rho_2^{\rm
      reference}(Q)} =
  \frac{\sav{\hro(Q^{aa})}}{\sav{\sav{\hro(Q^{ab})}}}
\end{align}
is a function of two entirely different quantities: the four-momentum
differences $Q^{aa}$ of ``sibling'' tracks taken from the same event
$a$ and one constructed from the mixed-event sample using tracks from
different events, written as $Q^{ab}$, $Q^{bc}$ etc. For second-order
correlations, the parent distribution is therefore necessarily a
two-variable probability\footnote{ We could argue that there are three
  different variables $Q^{aa}$, $Q^{ab}$ and $Q^{bc}$, where the last
  two differ in the sense that $Q^{ab}$ contains a track from the
  ``current'' event while $Q^{bc}$ does not. As shown below, this
  distinction is unnecessary as long as we keep careful track of
  possible occurrences of equal event indices.} %
$P(Q^{aa},Q^{bc})$ which, depending on whether the cases $b{=}a$ and
$c{=}a$ occur, may or may not factorise into a ``sibling'' and a
``mixed'' marginal probability
\begin{align}
  \lleq{vrf}
  P(Q^{aa},Q^{bc}) &= P_s(Q^{aa})\;P_m(Q^{bc})\quad\text{iff } a{\neq} b{\neq} c,
\end{align}
but (unless $a{=}b{=}c$) the marginals will always be
\begin{align}
  \lleq{vrfa}
  P_s(Q^{aa}) &= \int dQ^{bc}\,P(Q^{aa},Q^{bc}), \\
  P_m(Q^{bc}) &= \int dQ^{aa}\,P(Q^{aa},Q^{bc}).
\end{align}
The shapes of $P_s(Q)$ and $P_m(Q)$ must necessarily be different
since it is precisely this difference that leads to a nontrivial
signal in (\ref{vre}).  In terms of this joint probability, we can
write expectation values of eventwise counters (separately for
inclusive, fixed-$N$  or fixed-$n$ cases) as 
\begin{align}
  \lleq{vrg}
  E[\hro(Q^{aa})]
  &= \int_{\Omega} dQ^{aa}dQ^{bc}\;P(Q^{aa},Q^{bc})\,\hro_{aa}
  = \sum_{i\ne j} \int_{\Omega} dQ^{aa}\;P_s(Q^{aa})\,\delta(Q^{aa}-Q_{ij}^{aa})
  = \sum_{i\ne j} P_s(Q_{ij}^{aa}),
  \\
  \lleq{vrh}
  E[\hro(Q^{bc})] %
  &= \int_{\Omega} dQ^{aa}dQ^{bc}\;P(Q^{aa},Q^{bc})\,\hro_{bc}
  = \sum_{i,j} \int_{\Omega} dQ^{bc}\;P_m(Q^{bc})\,\delta(Q^{bc}-Q_{ij}^{bc})
  = \sum_{i,j} P_m(Q_{ij}^{bc}).
\end{align}
Later, we shall meet expectation values for cases such as $a{=}c$,
\begin{align}
  \lleq{vri}
  E[\hro(Q^{aa})\,\hro(Q^{ab})]
  &= \sum_{i\ne j}\sum_{k,\ell} \int_\Omega dQ^{aa}\,dQ^{ab}\;P(Q^{aa},Q^{ab})\;
  \delta(Q^{aa}-Q_{ij}^{aa})
  \delta(Q^{ab}-Q_{k\ell}^{ab})\nonumber\\
  &= \sum_{i\ne j}\sum_{k,\ell} P(Q_{ij}^{aa},Q_{k\ell}^{ab})
\end{align}
which definitely does not factorise.  The above expressions can be
simplified because we know that the parent distribution is not a
function of the individual track indices $i,j,k,\ell$
\begin{align}
  \lleq{vrj}
  P(Q_{ij}^{aa},Q_{k\ell}^{bc}) &= P(Q^{aa},Q^{bc})\qquad \forall\; i,j,k,\ell
\end{align}
and similarly $P_s(Q_{ij}^{aa}) = P_s(Q^{aa})$ and
$P_m(Q_{k\ell}^{bc})=P_m(Q^{bc})$.
For the event-averaged counters, this results in
\begin{align}
  \lleq{vrl}
  E[\rho_{aa}(Q)] &= \savN{n_a^{\ff{2}}} P_s(Q^{aa}) \\
  E[\rho_{bc}(Q)] &= \savN{n_b}\savN{n_c} P_m(Q^{bc})
\end{align}
or in terms of the notation %
of Section \ref{sec:qrs},
\begin{align}
  \lleq{vrm}
  E[\rhon{2}(Q)] &= \savN{n^{\ff{2}}} P_s(Q^{aa}) \\
  E[\rhon{1}{\otimes}\rhon{1}(Q)] &= \savN{n}^2 P_m(Q^{bc}).
\end{align}
As mentioned, we do not need the factorisation (\ref{vrf}) of
$P(Q^{aa},Q^{bc})$ as long as we keep careful track of the
equal-event-indices cases. Whenever $a{\neq}b$ or $a{\neq}c$,
independence of the events ensures that expectation values of products
of any functions $f(Q^{aa})$ and $g(Q^{ab})$ of the pair variables do
factorise,
\begin{align}
  \lleq{vrn}
  E[f(Q^{aa})\,g(Q^{ab})] &=  E[f(Q^{aa})]\,E[g(Q^{ab})] \qquad a\neq b
\end{align}
For third-order correlations, the parent distribution is a function of
three different variables $Q^{aaa}$, $Q^{bbc}$ and $Q^{def}$
containing respectively three, two or one track from the same event
and corresponding considerations regarding equal and unequal event indices
apply there, too.

\subsection{Statistical error calculation from first principles}
\label{sec:frqa}

It was shown in Section \ref{sec:exct} that expectation values would
have well-defined meanings in terms of underlying parent distributions
and their marginals \textit{if} their parent distributions were known,
which, however, they are not. We are therefore forced to revert from
expectation values $E[\cdot]$ to sample averages $\langle\cdot\rangle$
after completing a calculation.  The real use of such expectation
values in frequentist statistics has been in the form of a
gedankenexperiment which we now reproduce from Kendall
\cite{Ken86a}. Let $x$ be any generic eventwise counter or any other
eventwise statistic. Since the formulae in this section remain true
for inclusive and fixed-$N$ samples, we omit any notation related to
$N$ in this derivation.  In this simplified notation, the well-known
standard error of the sample mean $\sav{x}$ is given by (simplifying
$\nev-1 \to \nev$)
\begin{align}
  \lleq{ssk}
  \sigma(\sav{x}) 
  &= \sqrt{\var(\sav{x})}
  \ =\ \sqrt{\tfrac{1}{\nev}{\left[\sav{x^2}-\sav{x}^2\right]}}
\end{align}
which follows from the combinatorics of equal and unequal event
indices by the above artificial use of expectation values, reverting
from expectation values $E[\cdot]$ to sample means $\sav{\cdot}$ in
the last step:
\begin{align}
  \lleq{sska}
  \var(\sav{x}) 
  &= E[\sav{x}^2] - E[\sav{x}]^2 
  = \frac{1}{\nev^2} \sum_{a,b} \Bigl[E[x_ax_b] - E[x_a]\,E[x_b] \Bigr]
  \\
  &= \frac{1}{\nev^2} \sum_{a=b} \Bigl[ E[x_ax_b] - E[x_a]\,E[x_b] \Bigr]
  + \frac{1}{\nev^2} \sum_{a\ne b} \Bigl[ E[x_a]\,E[x_b] - E[x_a]\,E[x_b] \Bigr]
  \nonumber\\
  \lleq{sskb}
  &= \frac{1}{\nev^2} \sum_a \Bigl[ E[x^2] - E[x]^2 \Bigr] + 0
  = \frac{1}{\nev} \Bigl[ \sav{x^2} - \sav{x}^2 \Bigr],
\end{align}
where we have used the fact that $E[x_ax_b] = E[x_a]\,E[x_b]\;\forall
a{\ne}b$ and assumed that all $x$ are identically distributed, $E[x_a]
= E[x]\;\forall a$. Equality or inequality of event indices is thus
crucial. We shall follow the same approach below, keeping careful
track of equal and unequal event indices, factorising expectation
values for unequal event indices, and reverting to sample means in the
last step.

\subsection{Variances and covariances for multiple event averages}
\label{sec:vcov}

\subsubsection{Statistical errors for second-order cumulants}
\label{sec:sesc}

According to Eqs.~(\ref{ssc})--(\ref{ssi}), we must handle variances
and covariances of products of several event averages.  To derive
these, we shall use the following shortened notation: Letting
$\delta_{ij}^{ab} \equiv \delta(Q-Q_{ij}^{ab})$ etc, then $\hro_{aa} =
\sum_{i\ne j} \delta(Q-Q_{ij}^{aa}) = \sum_{i\ne j}\delta_{ij}^{aa}$
is the eventwise pair counter for event $a$, while $\hro_{bc} =
\sum_{i,j}\delta(Q-Q_{ij}^{bc})= \sum_{i,j}\delta_{ij}^{bc}$ is the
mixed-event counter of events $b$ and $c$ (with $b{\ne} c {\ne} a$
assumed), so that $\rho_2(Q) = \savo{\hro_{aa}}{a}$ while
$\rho_1{\otimes}\rho_1(Q) = \savo{\hro_{ab}}{ab}$ is a double event
average. We reserve the event index $a$ for the ``sibling'' event
whose correlations are currently being analysed, and use indices
$b,c,\ldots,t,u,v,w,\ldots$ for events entering the event-mixing
parts.\footnote{While it is irrelevant whether event $a$ is included
  or excluded in theoretical calculations of event mixing, it should
  never be used in actual implementations of mixing.} %
All quantities are assumed to be measured within a particular
subsample $\sampleN$ but we omit the $N$-subscript and the
argument. The event-index subscripts such as $\langle\cdot\rangle_{bc}$
above are included or omitted depending on whether they convey
relevant information on the specific averaging.

In this notation, the method that led to Eq.~(\ref{sskb}) reads
\begin{align}
  \lleq{scc}
  \var(\rho_2)
  &= \var(\savo{\hro_{aa}}{a})
  = \frac{1}{\nev^2} \sum_{a, b} \Bigl[
  E(\hro_{aa}\hro_{bb}) - E(\hro_{aa})\,E(\hro_{bb}) \Bigr]
  = \frac{1}{\nev}\Bigl[ \sav{(\hro_{aa})^2} - \sav{\hro_{aa}}^2 \Bigr].
\end{align}
The same method of disentangling the combinatorics of equal and
unequal event indices is applied consistently to all variances and
covariances below. The $\rho_1^2$-term in second-order cumulants
has variance
\begin{align}
  \lleq{scd}
  \var(\rho_1{\otimes}\rho_1)
  &= \var(\savo{\hro_{bc}}{bc})
  = \frac{1}{(\nev^{\ff{2}})^2} \sum_{b\ne c} \sum_{d\ne e} \Bigl[
  E(\hro_{bc}\hro_{de}) - E(\hro_{bc})\,E(\hro_{de})
  \Bigr].
\end{align}
The case $b{\ne}c{\ne}d{\ne}e$ yields zero, but the cases
$b{=}d{\ne}c{\ne}e$ and three other equivalent combinations yield
\begin{align}
  \lleq{sce}
  \var(\rho_1{\otimes}\rho_1)
  &= \frac{4\nev^\ff{3}} {(\nev^{\ff{2}})^2}
  \Bigl[ E[\hro_{bc}\hro_{be}] - E[\hro_{bc}]\,E[\hro_{be}] \Bigr] 
  \nonumber\\
  &\to\ %
  \frac{4\nev^\ff{3}} {(\nev^{\ff{2}})^2}
  \Bigl[ \savo{\hro_{bc}\hro_{be}}{bce} - \sav{\hro_{bc}}\,\sav{\hro_{be}} \Bigr]
  \quad \to \quad
  \frac{4} {\nev}
  \Bigl[ \sav{\hro_{bc}\hro_{be}} - (\rho_1{\otimes}\rho_1)^2 \Bigr]
\end{align}
where in the second step we reverted $E[\,]\to\sav{\,}$ and in the
third\footnote{Due to the factorisation of the expectation values
  earlier on, the fact that index $b$ appears in two separate sample
  averages does \textit{not} prevent us from replacing
  $\sav{\hro_{bc}}$ and $\sav{\hro_{be}}$ by
  $(\rho_1{\otimes}\rho_1)^2$.}  %
assumed $\nev\gg 1$. Note that the requirement $b{\ne}c{\ne}e$ implies
that
$\savo{\hro_{bc}\hro_{be}}{bce} %
= \savo{\savo{\sum_{i\ne j}\delta_{ij}^{bc}}{c} %
  \savo{\sum_{k\ne l} \delta_{k\ell}^{be}}{e}}{b} $ %
cannot be simplified to the square of a single counter %
$\savo{ \bigl\langle\sum_{i\ne
    j}\delta_{ij}^{bc}\bigr\rangle_c^2}{b}$: %
the event mixing involves three different events, not two.
Secondly, the combinations of two equalities $b{=}d{\ne} c{=}e$ and
$b{=}e{\ne} c{=}d$ in (\ref{scd}) yield another term of order
$\nev^{-2}$,
\begin{align}
  \lleq{sceb}
  \frac{2}{\nev^\ff{2}}
  \Bigl[ \savo{\hro_{bd}\hro_{bd}}{bd} - (\rho_1{\otimes}\rho_1)^2 \Bigr]
\end{align}
which we can safely neglect when $\sav{\np^2}/\nev\ll 1$ except when
there are few bins or large multiplicities even in small bins. %
It is worth emphasising that the extra factor 4 which appears in
Eq.~(\ref{sce}) arises from the same method that has been used for
decades to justify use of Eq.~(\ref{ssk}). We find, by the same
method, that the covariance between $\rho_2$ and
$\rho_1{\otimes}\rho_1$ is given by
\begin{align}
  \lleq{scf}
  \cov(\rho_2,\rho_1{\otimes}\rho_1)
  &= \cov(\sav{\hro_{dd}},\sav{\hro_{bc}})
  = \frac{1}{\nev\,\nev^{\ff{2}}} 
  \sum_d \sum_{b\ne c} \Bigl[ E(\hro_{dd}\hro_{bc}) - E(\hro_{dd})\,E(\hro_{bc})\Bigr]
  \\
  \lleq{scg}
  &= \frac{2} {\nev}
  \Bigl[ \sav{\hro_{dd}\hro_{db}} - \sav{\hro_{dd}}\sav{\hro_{dc}}\Bigr]
  = \frac{2} {\nev}
  \Bigl[ \sav{\hro_{dd}\hro_{dc}} - (\rho_2)(\rho_1{\otimes}\rho_1)\Bigr]
\end{align}
so that we must in addition accumulate, for every event $d$, the
product of the counters
\begin{align}
  \lleq{sch}
  \hro_{dd}\hro_{dc} = \sum_{i\ne j} \delta_{ij}^{dd} %
  \sum_k  \Bigl\langle\sum_\ell\delta_{k\ell}^{dc}\Bigr\rangle_c.
\end{align}
Note that there is no restriction on track indices $k{\ne}i$ or
$k{\ne}j$ in the $d$-event, meaning that events with $n(a) =2$
contribute to this counter which would otherwise not be the case.
Combining these, we find, to leading order in $\nev^{-1}$ and renaming
mixed-event indices,
\begin{align}
  \lleq{schb}
  \var(\kappa_2^I)
  &= \frac{1}{\nev} \left\{
    \sav{ 
      \left( \hro_{aa} - 2 F_{2\smallN} \hro_{ac} \right)
      \left( \hro_{aa} - 2 F_{2\smallN} \hro_{ad} \right) }_a
    - \left(\rho_2 - 2F_{2\smallN}\,\rho_1{\otimes}\rho_1 \right)^2
  \right\} 
\end{align}
with all event indices strictly unequal and $c$- and $d$-event
averages understood where appropriate.%
\footnote{While this factorised form is instructive, it cannot be used
  directly since $F_{2\smallN}$ can be determined only on completion
  of the entire sample analysis. Each of the counter products in
  (\ref{schb}) must hence be implemented separately.} %
Contrasting this with the traditional way to calculate the same
variance,
\begin{align}
  \lleq{schc}
  \var(\kappa_2^I)
  &= \frac{1}{\nev} \left\{
    \sav{ 
      \left( \hro_{aa} - F_{2\smallN} \hro_{ac} \right)
      \left( \hro_{aa} - F_{2\smallN} \hro_{ad} \right) }
    - \left(\rho_2 - F_{2\smallN}\,\rho_1{\otimes}\rho_1 \right)^2
  \right\} 
\end{align}
it is clear that in previous analyses the two possible ways to set $a$
equal to $b$ or $c$ were overlooked, while normal (non-internal)
cumulants also omit the $F_{2\smallN}$.

\subsubsection{Statistical errors for third-order cumulants}
\label{sec:setc}

In third order, we shall need $\delta_{ijk}^{abc} \equiv
\delta(\qghpa-Q_{ijk}^{abc})$ and similar quantities, and the notation
for counters $\hro_{aaa}$, $\hro_{aab}$ and $\hro_{abc}$ corresponding
to the event averages $\rho_3(\qghpa) = \sav{\hro_{aaa}}$,
$\rho_2{\otimes}\rho_1(\qghpa) = \sav{\hro_{aab}}$ and
$\rho_1{\otimes}\rho_1{\otimes}\rho_1(\qghpa) = \sav{\hro_{abc}}$
respectively. Clearly, $a{\ne}b{\ne}c$ must hold in the third order
case. We obtain for the necessary third-order quantities (shuffling
and/or renaming indices if necessary)
\begin{align}
  \lleq{sci}
  \var(\rho_3)
  &= \frac{1}{\nev}\Bigl[ \savo{(\hro_{aaa})^2}{a} - \rho_3^2 \Bigr]
  \\
  \lleq{scj}
  \cov(\rho_3,\rho_2{\otimes}\rho_1)
  &= \frac{1}{\nev\,\nev^{\ff{2}}} \sum_r \sum_{s\ne t}
  \Bigl[ E(\hro_{rrr}\hro_{sst}) - E(\hro_{rrr})\,E(\hro_{sst})\Bigr]
  \nonumber\\
  &= \frac{1}{\nev}   \Bigl[
  \sav{\hro_{rrr}\hro_{rrs}} + \sav{\hro_{rrr}\hro_{rss}}
  - 2\rho_3 (\rho_2{\otimes}\rho_1) \Bigr]
\end{align}
with
  $\hro_{rrr}\hro_{rrs}
  = \sum_{i\ne j\ne k} \delta_{ijk}^{rrr}
  \sum_{\ell\ne m}\Bigl\langle\sum_n\delta_{\ell mn}^{rrs}\Bigr\rangle_s$
and  
  $\hro_{rrr}\hro_{rss}
  = \sum_{i\ne j\ne k} \delta_{ijk}^{rrr}
  \sum_{\ell} \Bigl \langle \sum_{m\ne n} \delta_{\ell mn}^{rss} \Bigr\rangle_s$.
The remaining variances and covariances needed for third-order
correlations with GHP topology are, after renaming of indices,
\begin{align}
  \lleq{scka}
  \var(\rho_2{\otimes}\rho_1)
  &= \frac{\nev^{\ff{3}}}{(\nev^{\ff{2}})^2} \sum_{g\ne e}\sum_{c\ne d}
   \Bigl[E(\hro_{gge}\hro_{ccd}) - E(\hro_{gge})\,E(\hro_{ccd}) \Bigr]
   \nonumber\\
   &= \frac{\nev^{\ff{3}}} {(\nev^{\ff{2}})^2} \Bigl[
     \sav{\hro_{ggd}[\hro_{ggc} + \hro_{gcc} + \hro_{ddc} + \hro_{dcc}]}
   - 4(\rho_2{\otimes}\rho_1)^2 \Bigr],
\end{align}
while we neglect
\begin{align}
  \lleq{sckb}
  \frac{1}{\nev^\ff{2}} \Bigl[
  \hro_{ggc}\hro_{ggc} + \hro_{ggc}\hro_{gcc} - 2 (\rho_2{\otimes}\rho_1)^2  \Bigr].
\end{align}
The next term is simpler,
\begin{align}
  \lleq{scl}
  \cov(\rho_3,\rho_1^3)
  &= \frac{1}{\nev\,\nev^{\ff{3}}} \sum_t \sum_{u\ne v\ne w} \Bigl[
  E(\hro_{ttt}\hro_{uvw}) - E(\hro_{ttt})\,E(\hro_{uvw}) \Bigr]
  = \frac{3}{\nev} \Bigl[ \sav{\hro_{ttt}\hro_{tuv}}
  - \rho_3 \rho_1^3 \Bigr],
\end{align}
but the following is not,
\begin{align}
  \lleq{scm}
  \cov(\rho_2{\otimes}\rho_1,\rho_1^3)
  &= \frac{1}{\nev^{\ff{2}}\nev^{\ff{3}}} \sum_{u\ne v}\sum_{x\ne y\ne z} \Bigl[
  E(\hro_{uuv}\hro_{xyz}) - E(\hro_{uuv})\,E(\hro_{xyz}) \Bigr]
  \\
  \lleq{scn}
  &= \frac{3\nev^{\ff{4}}}{\nev^{\ff{2}}\nev^{\ff{3}}} \Bigl[ \sav{\hro_{wwx}\hro_{wyz}}
  + \sav{\hro_{wxx}\hro_{wyz}} - 2(\rho_2\rho_1)(\rho_1^3) \Bigr]
    + \frac{6}{\nev^{\ff{2}}} \Bigl[ \sav{\hro_{wwx}\hro_{wxy}}
    - (\rho_2\rho_1)(\rho_1^3) \Bigr],
\end{align}
and the large number of combinations makes the variance of $\rho_1^3$ particularly
complicated,
\begin{align}
  \lleq{sco}
  \var(\rho_1{\otimes}\rho_1{\otimes}\rho_1)
  &= \frac{1}{(\nev^{\ff{3}})^2}  \sum_{d\ne e\ne h} \sum_{b\ne p \ne q}
  \Bigl[ E(\hro_{deh}\hro_{bpq}) - E(\hro_{deh})\,E(\hro_{bpq}) \Bigr]
  \\
  &= \frac{9\nev^{\ff{5}}}{(\nev^{\ff{3}})^2} \Bigl[
     \savo{\hro_{beh}\hro_{bpq}}{} - (\rho_1^3)(\rho_1^3) \Bigr]
  + \frac{18\nev^{\ff{4}}}{(\nev^{\ff{3}})^2} \Bigl[
     \savo{\hro_{beh}\hro_{beq}}{} - (\rho_1^3)(\rho_1^3) \Bigr]
     \nonumber\\
  \lleq{scp}
  &+ \frac{6}{\nev^{\ff{3}}} \Bigl[
     \savo{\hro_{beh}\hro_{beh}}{} - (\rho_1^3)(\rho_1^3) \Bigr].
\end{align}
For large $\nev$, the leading order terms will usually dominate, so
that we can neglect the subleading terms.\footnote{%
  If and when large bins are used and the sixth power of the measured
  positive-pion multiplicity becomes comparable to $\nev$, subleading
  terms will have to be included.
  This requirement is less trivial than it may
  sound, since for subsamples of fixed multiplicity $N$, the number of
  events $\nevN$ is much smaller than $\nev$, while of course $\np$
  may be substantial when $N$ is large. For UA1, $\nevN =O(10^4)$
  while $\nev = O(10^6)$.} %
To leading order, we therefore obtain after substitution in
(\ref{ssf}) and again omitting brackets for non-$a$ event averages
\begin{align}
  \lleq{scpa}
  \var(\kappa_3^I) 
  &= \frac{1}{\nev} \biggl\{
    \biggl\langle
      \hro_{aaa}^2 + 9 \hro_{aab}( \hro_{aac}+\hro_{acc}+\hro_{bbc}+\hro_{bcc})
      + 9 G_3^2\hro_{aaa}(\hro_{abc}\hro_{ade})
      \\
      &\qquad\qquad 
      - 6 \hro_{aaa}(\hro_{aab}+\hro_{abb})
      + 6G_3 \hro_{aaa}\hro_{abc}
      - 18 G_3 \hro_{aab}(\hro_{acd}+\hro_{bcd})
      \biggr\rangle_{\!\!a}
      \nonumber\\
      &\qquad - \biggl[ 
        \rho_3^2 + 36 (\rho_2{\otimes}\rho_1)^2 + 9 G_3^2\,(\rho_1^3)^2
        - 12\rho_3(\rho_2{\otimes}\rho_1)
        + 6G_3\,\rho_3(\rho_1)^3 - 36G_3^2(\rho_2{\otimes}\rho_1)(\rho_1)^3
      \biggr]
    \biggr\} \nonumber\\
    \lleq{scq}
  &= \frac{1}{\nev} \biggl\{
    \biggl\langle
    ( \hro_{aaa} - 3\hro_{aab} - 3\hro_{abb} + 3G_3\hro_{abc} )
    ( \hro_{aaa} - 3\hro_{aad} - 3\hro_{add} + 3G_3\hro_{ade} )
    \biggr\rangle_{\!\!a}
    \nonumber\\
      &\qquad 
      - \biggl[\rho_3 - 6 \rho_2{\otimes}\rho_1 + 3G_3(\rho_1)^3  \biggr]^2
    \biggr\}.
\end{align}
While the factorised form is again instructive, it cannot be
calculated in this form within the $a$-loop in the analysis since the
$G_{3\smallN}$ constants are known only on completion of the entire
sample analysis. Rather, the full palette of product counters
$\hro\hro$ has to be accumulated and averaged and combined only in the
final phase of the analysis.

\section{Averaged internal cumulants} 
\label{sec:smpab}

As $N$ is only an approximation for the true total event multiplicity
anyway, and for cases of small sample statistics, it may be necessary
or desirable to group subsamples of fixed $N$ into multiplicity
classes $N\in[A,B]$. It is important, however, not to simply lump all
events within this multiplicity class into a single ``half-inclusive''
subsample, because, as has long been known
\cite{1975-Foa-pr22.1-xx.xx}, that results in terms entering the
cumulants which arise solely to ``multiplicity mixing'' (MM) of events of
different $N$. Given the arbitrary choice of $[A,B]$, such MM
correlations are spurious and avoided in favour of
``Averaged-Internal'' (AI) correlations\footnote{Historically, this
  issue was discussed under the name ``Short-Range Correlations'' and
  ``Long-Range Correlations'' \cite{1975-Foa-pr22.1-xx.xx}. Since
  current usage of the term ``Short-Range Correlations'' refers to
  correlations over small scales in momentum space, we rather define
  them more accurately as ``Averaged Internal'' (AI) correlations and
  ``Multiplicity-Mixing'' (MM) correlations, noting also that the
  correction factors in Eqs.~(\ref{nvd})--(\ref{nvj}) do not appear in
  the earlier literature.} %
defined as follows. Using the renormalised multiplicity distribution
\begin{align}
  \lleq{nvc}
  \evrNp 
  = \frac{\evrN}{\sum_{N=A}^B \evrN}
  = \frac{\nevN}{\sum_{N=A}^B \nevN}
\end{align}
the AI unnormalised cumulants, reference distributions and normalised
AI cumulants are %
\begin{align}
  \lleq{nvd}
  \kappa_2^{AI}(Q\cond \sampleAB) 
  &= \sum_{N=A}^B \evrNp \;\kappa_2^I(Q\cond \sampleN),
  \\
  \lleq{nve}
  \rho_1{\otimes}\rho_1(Q\cond \sampleAB) 
  &= \sum_{N=A}^B \evrNp \;F_{2\smallN}\;\rho_1{\otimes}\rho_1(Q\cond \sampleN),
  \\
  \lleq{nvf}
  K_2^{AI}(Q\cond \sampleAB) 
  &= \frac{\kappa_2^{AI}(Q\cond \sampleAB)}{\rho_1{\otimes}\rho_1(Q\cond \sampleAB)} 
  \ =\ \frac{\sum_{N=A}^B \evrNp \,\rho_2(Q\cond \sampleN)} %
  {\sum_{N=A}^B \evrNp \,F_{2\smallN}\,\rho_1{\otimes}\rho_1(Q\cond \sampleN)}
  \ -\ 1.
\end{align}
Note that the correction factors $F_{2\smallN}$, which are normalised
factorial moments of $n$ for fixed $N$, are part of the summed
normalisations.%
\footnote{ %
  An expression with a correction factor outside the sums such as
  \begin{align}
    K_2 &= \frac{\sav{\np}^2}{\sav{\np(\np-1)}}\cdot
    \frac{\sum_N \evrNp \kappa_2^I(Q\cond \sampleN)}%
    {\sum_N \evrNp \rho_1{\otimes}\rho_1(Q\cond \sampleN)} 
    \nonumber
  \end{align}
  is inconsistent with AI correlation averaging if the single-particle
  spectra or some other physical effect change significantly within
  the range $[A,B]$.  We also note that the above differs from the
  formula used in
  Ref.~\cite{2006-EggersOctoberBuschbeck-plb635.280-hepex0601039} for
  second-order correlations in $\bmq$.  In the present notation, the
  cumulant used in
  Ref.~\cite{2006-EggersOctoberBuschbeck-plb635.280-hepex0601039}
  reads
  \begin{align}
    K_2(\bmq) &= 
    \frac{\sum_N \evrNp \rho_2(\bmq\cond \sampleN)} %
    {\sum_N \evrNp \tfrac{N-1}{N} \rho_1{\otimes}\rho_1(\bmq\cond \sampleN)}
    \ -\ 1 \nonumber
  \end{align}
  i.e.\ a correction for an $N$-multinomial rather than the weighted
  sum of $n$-multinomials used in
  Eqs.~(\ref{nvd})--(\ref{nvf}). } %
In third order, we have correspondingly
\begin{align}
  \lleq{nvh}
  \kappa_3^{AI}(\qghpa\cond \sampleAB) 
  &= \sum_{N=A}^B \evrNp \;\kappa_3^I(\qghpa\cond \sampleN),
  \\
  \lleq{nvi}
  \rho_1{\otimes}\rho_1{\otimes}\rho_1(\qghpa \cond \sampleAB) 
  &= \sum_{N=A}^B \evrNp \;F_{3\smallN}\;\rho_1{\otimes}\rho_1{\otimes}\rho_1(\qghpa\cond \sampleN),
  \\
  \lleq{nvj}
  K_3^{AI}(\qghpa\cond \sampleAB) 
  &= \frac{\kappa_3^{AI}(\qghpa\cond \sampleAB)}
  {\rho_1{\otimes}\rho_1{\otimes}\rho_1(\qghpa\cond \sampleAB)}.
\end{align}
Note also that (\ref{nvi}) holds for the normalisation only and not
for the last term in $\kappa_3^I$, which is
$(3F_{2\smallN}-F_{3\smallN})\;\rho_1{\otimes}\rho_1{\otimes}\rho_1$ --- but that
is already taken care of in the formula (\ref{kqg}) for $\kappa_3^I$
itself.
Expressions for an inclusive (all-$N$) multiplicity summation of
internal cumulants are obtained from the above by setting $A{=}0$ and
$B=\infty$.
The AI (Averaged Internal) correlations Eqs.~(\ref{nvd}) and
(\ref{nvh}) represent refined versions of what has traditionally been
termed ``Short-Range Correlations'', differing from the original
formulae \cite{1975-Foa-pr22.1-xx.xx,1991-Carruthers-pra43.2632-xx.xx}
by the $F_{2\smallN}$ and $G_{3\smallN}$ factors respectively. This
was originally pointed out in
Ref.~\cite{1996-LipaEggersBuschbeck-prd53.4711-hepph9604373} but only
for multinomials in $N$.

Regarding variances and standard errors for AI correlations, we first
note that, since subsamples $\sampleN$ are strictly mutually
independent, a variance over the $[A,B]$ range is simply the weighted
sum of the corresponding fixed-$N$ variances.  From Eqs.~(\ref{nvd})
and (\ref{nvj}) we have to all orders $r$
\begin{align}
  \lleq{nvk}
  \kappa_r^{AI}(Q\cond \sampleAB) 
  &= \sum_{N=A}^B \evrNp \;\kappa_r^I(Q\cond \sampleN) \qquad r=2,3,\ldots\,,
\end{align}
and given the independence of any functions $f$ and $g$ of different
multiplicity subsamples, $E[f(\sampleN)\cdot g({\mathcal
  S}_{\scriptscriptstyle N'})] = E[f(\sampleN)]\cdot E[g({\mathcal
  S}_{\scriptscriptstyle N'})]\; \forall\;N\neq N'$, we conclude that
\begin{align}
  \lleq{nvl}
  \var[\kappa_r^{AI}(Q\cond \sampleAB)]
  &= \sum_N (\evrNp)^2 \;\var[\kappa_r^I(Q\cond \sampleN)]
  \\
   \lleq{nvm}
   \var[\rho_1^r(Q\cond \sampleAB)]
   &= \sum_N (\evrNp)^2 \;(F_{r\smallN})^2\;\var[\rho_1^r(Q\cond \sampleN)]
   \\
  \lleq{nvn}
  \cov[\kappa_r^{AI}(Q\cond \sampleAB),\rho_1^r(Q\cond\sampleAB)]
  &= \sum_N (\evrNp)^2 \;F_{r\smallN}\;\cov[\kappa_r^I(Q\cond\sampleN),\rho_1^r(Q\cond\sampleN)]
\end{align}
which are known functions in terms of Sections \ref{sec:sesc} and
\ref{sec:setc}, while for the normalised cumulants in $[A,B]$,
covariances between numerator and denominator are (omitting the $Q$),
\begin{align}
  \lleq{nvp}
  \cov[\kappa_2^I(\sampleN), \rho_1^2(\sampleN)]
  &= \cov[\rho_2(\sampleN),\rho_1{\otimes}\rho_1(\sampleN)] 
  - F_{2\smallN} \var(\rho_1{\otimes}\rho_1(\sampleN))
  \\
  \lleq{nvq}
  \cov[\kappa_3^I(\sampleN), \rho_1^3(\sampleN)]
  &= \cov[\rho_3(\sampleN),\rho_1^3(\sampleN)]
    - 3\cov[\rho_2{\otimes}\rho_1(\sampleN),\rho_1^3(\sampleN)]
    + G_3\var[\rho_1^3(\sampleN)]
\end{align}
so that the normalised range cumulants have variances
\begin{align}
  \lleq{nvr}
  \var[K_r^{AI}(Q\cond \sampleAB)]
  &= (K_r^{AI})^2\cdot \Biggl[ 
  \frac{\var[\kappa_r(\sampleAB)]}{(\kappa_r(\sampleAB))^2} 
  + \frac{\var[\rho_1^r(\sampleAB)] }{(\rho_1^r(\sampleAB))^2}
  - 2\frac{\cov[\kappa_r(\sampleAB),\rho_1^r(\sampleAB)]} {\kappa_r(\sampleAB)\cdot\rho_1^r(\sampleAB)}
  \Biggr]
\end{align}
and standard errors are given by\footnote{%
  One might expect $\sigma(K_r^{AI}(\sampleAB))$ to include a
  prefactor of the sort seen in Eq.~(\ref{ssk}) i.e. something like
  $\sqrt{[\var(K_r^{AI}(Q\cond \sampleAB)]/[B-A]} $, but this would be
  incorrect.  The reason in that the formulae (\ref{nvl})--(\ref{nvn})
  for range $AB$ can be considered as an average, so that we can apply
  the methods of Section \ref{sec:frqa} to obtain the same
  results. For example, considering $\kappa_2^{AI}(Q\cond \sampleAB)
  \equiv \overline{\kappa_2}$ of (\ref{nvd}) as an average and writing
  $\kappa_2^I(Q\cond \sampleN) \equiv \kappa_{2\smallN}$, the variance
  on this average is
  \begin{align}
    \lleq{nvlb} \var[\overline{\kappa_2}]
    &= E[(\overline{\kappa_2})^2] - E[\overline{\kappa_2}]^2  \nonumber\\
    &= E[(\textstyle\sum_N \evrNp \kappa_{2\smallN})
    \cdot(\textstyle\sum_{N'} \evrNpp \kappa_{2\smallN'})]
    - E[(\textstyle\sum_N \evrNp \kappa_{2\smallN})]
    E[(\textstyle\sum_{N'} \evrNpp \kappa_{2\smallN'})]
    \nonumber\\
    &= \textstyle\sum_{N=N'} (\evrNp)^2 \left( E[\kappa_{2\smallN}^2] -
      E[\kappa_2^I(Q\cond\sampleN)]^2
    \right) 
    = \textstyle\sum_N (\evrNp)^2 \var(\kappa_{2\smallN}) \nonumber
  \end{align}
  which is identical with (\ref{nvl}). Therefore, division by $B-A$ is
  incorrect. %
}
\begin{align}
  \sigma(K_r^{AI}(\sampleAB)) &= \sqrt{\var(K_r^{AI}(Q\cond \sampleAB))}.
\end{align}

\section{Event mixing algorithms}
\label{sec:vtmx}

``Event mixing'' \cite{1974-Kopylov-plb50.472-xx.xx} is widely used to
simulate uncorrelated or semi-correlated quantities such as $\rho_1^3$
and $\rho_2{\otimes}\rho_1$. The idea has always been to use the
experimental sample at hand to simulate the baseline of independence
referred to in Section \ref{sec:ctra} in such a way that criteria 2
(independence of momenta), 4 (reproducing the one-particle momentum
space distribution) and 6 (normalisation) are all addressed
simultaneously.  Ideally, all effects bar the desired correlation are
elegantly removed in this way.

For the internal cumulants and their variances and covariances derived
above, event mixing requires keeping track of counters of all orders
in each of the subsamples $\sampleN$. A count of event indices in
Section \ref{sec:stderrors} shows that in a brute-force calculation we
would need, for each subsample, a minimum of five independent event
averages or $O(\nev^5)$ event combinations; furthermore, caution would
advise not to use the same event in calculating related counters, so
that selection and use of more than five events in mixing is
advisable. The resulting excessive number of full event averages,
mixing every (sub)sample event with every other one, is therefore not
feasible.

If the order of events in the sample is random, the multiple event
averages can be simplified by the use of the following multiple event
buffer algorithm.
\begin{enumerate}
\item A single overall event loop equivalent to the event index $a$
  runs over the entire inclusive sample $\samplei$. A given event $a$
  will have a multiplicity $N{=}N(a)$, so for that particular $a$
  correlation analysis for subsample $\sampleN$ is advanced by one
  event while the others remain dormant.\footnote{Inevitably, there
    are very few events in the high-multiplicity tail of the entire
    sample. These must be treated separately, e.g.\ by putting all
    events with $N$ greater than some threshold into a single
    buffer.} %
  In this way, $a$, which always refers to the sibling event, runs
  over all $\nevN$ events of every subsample $\sampleN$. There is no
  need to either explicitly sort the inclusive sample into subsamples
  or to run multiple jobs for fixed $N$.

\item The first $\nev_\smallB$ events\footnote{The number of events in
    a buffer $\nev_\smallB$ is usually kept the same for each buffer.}
  of a given multiplicity $N$ are used solely to fill up the buffer
  without doing any analysis. Once a given buffer has been filled,
  event mixing analysis proceeds for that subsample as follows:
  \begin{enumerate}

  \item An newly-read $a$-event is assigned to the $N{=}N(a)$ buffer,
    the earliest event in that buffer is discarded, and sums for
    averages entering $F_{2\smallN}$ and $F_{3\smallN}$ as well as the
    sibling counters $\hro_{aa}$, $\hro_{aa}$, $(\hro_{aa})^2$ and
    $(\hro_{aaa})^2$ are updated.

  \item Event combinations for mixed-event counters are built up by
    picking any one of the $\nev_\smallB-1$ other events in that
    buffer and calling it $b$, thereafter picking any one of the
    remaining $\nev_\smallB-2$ events in the buffer, calling it $c$
    and so on.  While to third order only five events (including the
    sibling event) are needed to construct all the counters required,
    in practice it is better to use different mixing events for
    different counters to root out even traces of unwanted
    correlations between different mixing counters.
    The random selection of events rather than tracks for mixing is
    necessary to ensure that more than one track per event can be used
    as required for counters of Sections
    \ref{sec:corrq}--\ref{sec:stderrors} such as $\hro_{bbc}$.

  \item For a given event set $b,c,d,\ldots$, the mixing counters are
    incremented using all possible combinations of the $\np_a$ tracks
    in event $a$ together with all the $\np_b, \np_c, \ldots$ tracks
    in the selected events $b,c,\ldots$ mixing events. For example
    $\hro_{bbc}$ would use all possible $\np_b(\np_b-1)$ pairs of
    $b$-tracks\footnote{As in the definition of the counters, each
      pair is counted twice: these are \textit{ordered pairs}.}
    together with all possible $\np_c$ single $c$-tracks. %
    The mixing of all tracks of a given event rather than just
    selected ones ensures that the fluctuations in $\np$ for given
    fixed $N$ are automatically contained in the counters.

  \item For constant $a$, the process of selecting events $b,c,\ldots$
    is repeated $C_{\rm mix} =$ 10--100 times to reduce the
    statistical errors, avoiding events that have been used in
    previous selections.
    Efficiency can be maximised by tuning of both the number of events
    $\nev_\smallB$ stored in each buffer and by the number of
    resamples $C_{\rm mix}$.

  \end{enumerate}
\item Once the entire sample has been processed via the $a$-event loop, the
  $b$- $c$- $d$-event averages are normalised by $C_{\rm
    mix}(\nevN-\nev_\smallB)$, while the primary $a$-average is
  normalised by $(\nevN-\nev_\smallB)$.

\item Unnormalised and normalised correlation quantities, their
 standard errors and correction factors are constructed by
  appropriate combinations of averaged counters.

\item Results from fixed-$N$ subsamples can then be combined into AI
  Correlations over partial ranges of $N$ or the entire inclusive
  sample at the end of the event loop using the methods outlined in
  Section \ref{sec:smpab}.
\end{enumerate}

\section{Discussion and Conclusions}
\label{sec:rdcs}

\begin{enumerate}
\item Correlations are only defined properly if the null case or
  reference distribution is defined on the same level of
  sophistication as the correlation itself. Translated into
  statistics, the six criteria set out in Section \ref{sec:ctra} for a
  reference sample for correlations at fixed charged multiplicity $N$
  lead straight to the definition of the reference sample as the
  average of multinomials given in Eq.~(\ref{mne}), weighted by the
  conditional multiplicity distribution $\evrnpN$. Assigning Bernoulli
  probabilities $\alpha(\bmp\cond \samplenpN) = \savN{\rho(\bmp\cond
    \samplenpN)}/ \savN{n}$ yields the reference density (\ref{mnk})
  and generating functional (\ref{mnea}).

\item From the theorem that internal cumulants are given by the
  difference between measured and reference cumulants we obtain
  normalised and unnormalised internal cumulants which satisfy every
  stated criterion for proper correlations for fixed-$N$ samples.

\item We have highlighted the distinction between $n$, the particles
  entering the correlation analysis itself, and $N$, the particles
  determining the event selection criterion for a particular
  semi-inclusive subsample. Various correction factors are shown to be
  fair to good approximations of these exact results in some cases but
  far off the mark in others.  Surprisingly, normalised cumulants are
  far more sensitive to these corrections, through the normalisation
  prefactor, than their unnormalised counterparts. To belabour the
  point: For any variables $(x_1,x_2)$, different definitions for
  correction factor $1/F$ for fixed-$N$ correlations of positive
  pions,
  \begin{align}
    K_2(x_1,x_2\cond \sampleN) &= \frac{1}{F} \frac{\rho(x_1,x_2\cond
      \sampleN)} %
    {\rho(x_1\cond \sampleN)\,\rho(x_2\cond \sampleN)} \ -\ 1,
  \end{align}
  can be very important at low multiplicities, with $F=1$ (Poisson)
  being the worst approximation, $F = N(N{-}1)/N^2$ a fair one, and
  $\savN{n(n{-}1)}/\savN{n}^2$ the best. 

  For inclusive correlations, correction factors such as $\langle
  n\rangle^2 / \langle n(n{-}1)\rangle_{\rm incl}$ were proposed early
  on in Refs.~\cite{1979-GyulassyKauffmannWilson-prc20.2267-xx.xx,
    1988-Zajc-ConfHMP.xx-xx.xx} 
  in an approach based on probabilities rather than densities.
  Ref.~\cite{1999-MiskowiecVoloshin-hip9.283-nuclex9704006} %
  specifically calls the inclusion of these correction factors
  meaningless because the theory then requires that the emission
  function be identically zero. We note that the argument in all those
  references relates to inclusive samples, while for the samples of
  fixed $N$ considered in this paper the prefactor, which is
  \textit{an average of $n$ at fixed $N$}, is a necessity. Either way,
  the arbitrariness of the use or non-use of the prefactor has been
  eliminated here based solely on considerations related to the
  reference distribution.

\item The problem posed in this paper, viz.\ the relation between
  charged multiplicity on the one hand and correlations based on the
  conditional $n$-distribution $\evrnpN$, has attracted little
  attention in the literature. Indeed, almost all theoretical work on
  multipion correlations, as for example summarised in
  Ref.~\cite{2001-HeinzScottoZhang-ap288.325-hepph0006150}, starts
  from the projection of final-state events, with all their different
  particle species, onto the single-species subspace of either $n_+$
  (positive pions) or $n_-$ (negative pions) correlations, to the
  exclusion of the other charge. It would be interesting to see a
  combined theory for both $n_+$ \textit{and} $n_-$, which would
  encompass all the work done so far plus correlations between
  unlike-sign pions and, of course, the issue raised by us here.

\item As shown in Figs.~2--3, correction factors for third order are
  larger than the second-order ones. For higher $r$-th order
  correlations, the effect of using a fixed-$N$ subsample is
  suppressed by approximately $1/\langle n^{r-1}\rangle_\smallN$ for
  unnormalised cumulants but actually worsens for normalised cumulants
  due to the normalisation prefactors $\langle n\rangle_\smallN^r /
  \langle n^{\ff{r}}\rangle_\smallN$ for small $n$. The importance of
  accurate correction of normalised quantities therefore rises with
  order of correlation.

\item The difference between poissonian and internal cumulants is
  largest at small multiplicities $n$. The mixed-multinomial
  prescription will therefore be required for \textit{any} correlation
  analysis involving small $n$, independently of the magnitude of
  $N$. Apart from the usual suspects of leptonic, hadronic and
  low-energy collisions, the low-$n$ case occurs both for very
  restricted phase space (such as in spectrometer experiments) and for
  correlations of rare particles such as kaons and baryons, even for
  large $N$.

\item Since $n$ fluctuates according to the conditional multiplicity
  distribution $\evrnpN$, the degree to which fixed-$N$ correlations
  differ from inclusive ones is strongly coupled to the character of
  $\evrnpN$. In general, $\evrnpN$ is subpoissonian and so
  $F_{r\smallN}$ falls below the poisson limit of 1.  The correction
  from fixed-$N$ poissonian to internal normalised cumulants is hence
  upward, not downward as in the case of multiplicity-mixing
  corrections.

\item While not the main subject of the present paper, some light is
  cast on the relationship between three levels of correlation, viz.\
  correlations inherent in the overall multiplicity distribution,
  multiplicity-mixing correlations, and the true internal correlations
  for fixed $N$. Each of these can and should be treated
  separately. The Averaged-Internal correlations of Section
  \ref{sec:smpab} are a compromise solution which may be useful both
  for physics reasons and for small datasets.

\item The fixed-$N$ corrections discussed here are separate and
  complementary to other important effects at low multiplicity.
  Refs.~\cite{2001-HeinzScottoZhang-ap288.325-hepph0006150,1987-Zajc-prd35.3396-xx.xx}
  highlight, for example, possible effects of ``residual
  correlations'' resulting from projecting from multipion to two-pion
  correlations.

  Energy-momentum conservation would also play a role.  Borghini
  \cite{2003-Borghini-epjc30.381-hepph0302139,
    2007-Borghini-prc75.021904-nuclth0612093} 
  has, for example, calculated the effect of momentum conservation for
  normalised two- and three-particle cumulants in momenta and
  $\bmp_t$. However, the saddlepoint method used applies to the
  large-$N$ limit and the results cannot be directly applied to the
  low-$N$ (and hence low-$n$) samples under discussion here.
  Indeed, momentum conservation will be near-irrelevant for cases of
  large $N$ and small $n$ as discussed above, but the small-$n$
  corrections of this paper will remain important. For the specific case of
  like-sign pion femtoscopy, the fact that only $n \sim \azia N/2
  \Omega$ out of the $N$ charged pions are used and that momentum
  conservation constraints include all other final-state particles
  both imply that momentum conservation constraints may be less
  important than the internal-cumulant correction introduced here.

  For the specific choice of correlation variables $Q$ and $Q_a$ for
  two- and three-particle cumulants, the contribution of momentum
  conservation to cumulants at small $Q$ will be small since the
  counts will be dominated by pairs at small $(\Delta\phi,\Delta y)$
  and intermediate $(p_{t1},p_{t2})$. %
  As pointed out by
  \cite{2008-ChajeckiLisa-prc78.064903-nuclth0803.0022}, momentum
  conservation exerts greatest influence at large pair or triplet
  momenta and hence mostly at large $Q$, where it may lead to moments
  and cumulants which do not converge to a constant as presupposed in
  most fits. The ad hoc method of multiplying fit parametrisations by
  a prefactor $1 + c \, Q$ with $c$ a free parameter does not
  adequately address the problem.

  Regarding the multiplicity dependence of the influence of
  energy-momentum conservation,
  Ref.~\cite{2009-ChajeckiLisa-prc79.034908-nuclth0807.3569}
  calculates the effect of energy-momentum conservation on
  single-particle differential observables, and finds significant
  systematic effects. No doubt this must also be the case for
  multiparticle observables, although, as we have pointed out above,
  the effect of conservation laws will be diluted by the fact that
  fewer than half of the final-state particles of any given event are
  actually used in the present analysis. Detailed investigations are
  beyond the scope of this paper.

\item We have also recalculated statistical errors for products of
  event averages starting from the original prescription which forms
  the basis of frequentist statistical error calculations. Compared to
  conventional statistical error calculations, new prefactors appear
  in our calculations --- see e.g.\ Eq.~(\ref{sce}) and the results in
  Section \ref{sec:setc} --- which have surprisingly been missed so
  far.

\item We note that the present formalism is still in the frequentist
  statistics mindset, which may be inaccurate for low multiplicities
  and should be supplanted by a proper Bayesian analysis. The final
  word has certainly not been spoken about correlation analysis of
  small-$n$ datasets.

\end{enumerate}

\subsection*{Acknowledgements}

This work was supported in part by the National Research Foundation of
South Africa.


\end{document}